\documentclass[lettersize,journal]{IEEEtran}
\IEEEoverridecommandlockouts 
\usepackage{cite}
\usepackage{float}  
\usepackage{subfigure}
\usepackage{booktabs} 
\usepackage[inkscapelatex=false]{svg}   
\usepackage{amsmath,amssymb,amsfonts}
\usepackage{algorithmicx}
\usepackage{graphicx}
\usepackage{textcomp}
\usepackage{xcolor}
\usepackage{soul, color, xcolor}
\usepackage{tikz,xcolor}
\usepackage{hyperref}
\usepackage[hyphenbreaks]{breakurl}
\usepackage{amsmath, amssymb, mathrsfs}
\usepackage{algorithm}
\usepackage{algpseudocode}
\usepackage{verbatim}
\usepackage{stfloats} 

\usepackage [most]{tcolorbox}
\usepackage{makecell}  
\tcbuselibrary{skins, breakable}
\tcbset{
    colback=gray!50 !white,
    colframe=gray! 75!black, 
    fonttitle=\bfseries, 
    arc=0pt, 
    boxrule=0.4pt, 
    title=my grey，
}
\definecolor{darkgreen}{RGB}{0,100,0}
\definecolor{OliveGreen}{RGB}{85,107,47}   
\definecolor{PineGreen}{RGB}{1,121,111}    
\definecolor{SoftGreen}{RGB}{46,139,87}
\definecolor{MutedGreen}{RGB}{80,120,60}  
\definecolor{SoftBlue}{RGB}{70, 130, 180} 
\definecolor{Cornflower}{RGB}{100, 149, 237} 

\def\BibTeX{{\rm B\kern-.05em{\sc i\kern-.025em b}\kern-.08em
    T\kern-.1667em\lower.7ex\hbox{E}\kern-.125emX}}

\definecolor{lime}{HTML}{A6CE39}
\DeclareRobustCommand{\orcidicon}{
\begin{tikzpicture}
\draw[lime, fill=lime] (0,0)
circle[radius=0.16]
node[white]{{\fontfamily{qag}\selectfont \tiny \.{I}D}};
\end{tikzpicture}
\hspace{-2mm}
}
\foreach \x in {A, ..., Z}{%
\expandafter\xdef\csname orcid\x\endcsname{\noexpand\href{https://orcid.org/\csname orcidauthor\x\endcsname}{\noexpand\orcidicon}}
}

\begin{document}
\title{CKM-Enabled Joint Spatial-Doppler Domain Clutter Suppression for Low-Altitude UAV ISAC}

\author{
Zihan~Xu, Zhiwen~Zhou, Di~Wu,~\IEEEmembership{Student Member,~IEEE}, Xiaoli~Xu,~\IEEEmembership{Member,~IEEE} and Yong~Zeng\,~\IEEEmembership{Fellow,~IEEE}

\thanks{An earlier version of this paper was presented in part at the 2024 IEEE/CIC International Conference on Communications in China (ICCC Workshops), Hangzhou, China, in August 2024 [DOI: 10.1109/ICCCWorkshops62562.2024.10693799].}
\thanks{The authors are with the National Mobile Communications
Research Laboratory, Southeast University, Nanjing 210096, China. Di Wu and Yong Zeng are also with the Purple Mountain Laboratories, Nanjing 211111, China (e-mail: \{zihanxu, zhiwen\_zhou, studywudi, xiaolixu, yong\_zeng\}@seu.edu.cn).\textit{(Corresponding author: Yong Zeng.)}}

}

\markboth{Journal of \LaTeX\ Class Files,~Vol.~14, No.~8, August~2021}%
{}

\maketitle

\begin{abstract} 
The rapid development of low-altitude economy has placed higher demands on the sensing of small-sized unmanned aerial vehicle (UAV) targets. However, the complex and dynamic low-altitude environment, like the urban and mountainous areas, makes clutter a significant factor affecting the sensing performance. Traditional clutter suppression methods based on Doppler difference or signal strength are inadequate for scenarios with dynamic clutter and slow-moving targets like low-altitude UAVs. In this paper, motivated by the concept of channel knowledge map (CKM), we propose a novel clutter suppression technique for orthogonal frequency division multiplexing (OFDM) integrated sensing and communication (ISAC) system, by leveraging a new type of CKM named clutter angle map (CLAM). CLAM is a site-specific database, containing location-specific primary clutter angles for the coverage area of the ISAC base station (BS). With CLAM, the sensing signal components corresponding to the clutter environment can be effectively removed before target detection and parameter estimation, which greatly enhances the sensing performance. Besides, to take into account the scenarios when the targets and clutters are in close directions so that pure CLAM-based spatial domain clutter suppression is no longer effective, we further propose a two-step CLAM-enabled joint spatial-Doppler domain clutter suppression algorithm. Simulation results demonstrate that the proposed technique effectively suppresses clutter and enhances target sensing performance, achieving accurate parameter estimation for sensing slow-moving low-altitude UAV targets.


\end{abstract}

\begin{IEEEkeywords}
Channel knowledge map, clutter suppression, low-altitude UAV, sensing parameters estimation, ISAC. 
\end{IEEEkeywords}

\section{Introduction}
\IEEEPARstart{L}{ow-altitude} unmanned aerial vehicles (UAVs) are crucial components to empower the emerging low-altitude economy, with a wide range of Internet of Things (IoT) applications like smart traffic, precision agriculture, aerial logistics and disaster rescue. However, for complex and dynamic low-altitude environments such as urban and mountainous areas, there exist significant challenges for the real-time and accurate sensing of low-altitude targets, especially those ‘low, slow, and small’ targets, due to the line-of-sight (LoS) blockage and strong clutter\cite{jiang2025integrated}\cite{zeng2019accessing}\cite{song2025overview}\cite{11165289}. These factors compromise the reliability of the entire low-altitude system.

Integrated sensing and communication (ISAC) has emerged as a promising technology in low-altitude environments\cite{dingyou2025integrated}. In fact, ISAC has been identified as one of the key usage scenarios for IMT-2030 (6G)\cite{ITU-R}. It enables communication and sensing to complement each other, enhancing sensing through information from the communication network\cite{liu2022integrated}. There have been significant research efforts on target sensing in ISAC systems, including information theory \cite{9705498}\cite{ahmadipour2023information}\cite{wang2024cramer}, waveform design\cite{xiao2022waveform}\cite{10771629}, beamforming\cite{mao2024communication}\cite{10659350}\cite{10382696}, and signal processing\cite{dai2025tutorial}\cite{wei2023integrated}\cite{zhao2024modeling}. However, research on clutter suppression for low-altitude ISAC in complex environments is relatively limited.

Clutter has a significant impact on the sensing performance of ISAC systems in complex low-altitude scenarios\cite{luo2024integrated}. In addition to the reflected/scattered signal from the target to the base station (BS), the presence of background scatterers such as trees and buildings in the low-altitude environment causes strong clutters. These clutters mix with the target signals at the receiver, severely interfering with target sensing and even causing the desired target signal to be overwhelmed in low signal-to-noise ratio (SNR) conditions. Therefore, enhancing the clutter suppression capability of ISAC systems in low-altitude clutter environments, and thereby achieving sensing enhancement, is urgently needed.

On the other hand, clutter suppression has been a research focus in the radar field, with a long history of study. The classic method is to distinguish between moving targets and static clutter based on their velocity difference. As the most commonly used technology, moving target indication (MTI) employs a high-pass Doppler filter\cite{mti}, placing the clutter signal within the stopband or transition band of the filter, thereby suppressing zero-Doppler and very low-Doppler clutter signals, while retaining the sensing signal of those targets with high Doppler. Furthermore, moving target detection (MTD) technology\cite{ludloff1985reliability}, based on MTI, uses a set of band-pass filters to screen targets with specific Dopplers, thereby attempting to more precisely estimate target velocity. Subsequent researchers proposed a series of methods to optimize the design of Doppler filters for different scenarios \cite{Optimum-MTI-design-2}\cite{short2007adaptive}\cite{wan2017improved}\cite{zhang2025longtime}. Nevertheless, the above methods rely on the assumption that the Dopplers between the clutter and targets are substantially different. The authors in \cite{luo2024integrated} tried to improve the ISAC performance in a cluttered environment, but it assumes that the clutter is static and the channel matrix of clutter can be precisely reconstructed. However, for slow-moving sensing targets such as low-altitude UAVs or when the clutter is dynamic, say due to the movement of the sensing transmitter/receiver, the difference of the Doppler between the sensing targets and clutter may be insignificant. In this case, the aforementioned Doppler-difference-based clutter suppression techniques become ineffective. 

Besides clutter suppression in the Doppler domain, classic radar also employs the clutter map technology to record the intensity of background clutter in the environment. Specifically, the physical environment is discretized into multiple range-angular cells for a moving window estimator, and the clutter detection threshold is determined for each cell through periodic accumulation, averaging, and updating \cite{7455429}. For example, \cite{nitzberg2007clutter}\cite{hamadouche2000analysis} focused on setting adaptive thresholds by scanning the background clutter power map cell by cell periodically, thereby enabling constant false alarm rate (CFAR) detection in non-uniform clutter backgrounds. \cite{xiangwei2010adaptive} introduced the ordered data variability (ODV) technique to address the heavy masking effects and reduce reliance on prior interference information. Furthermore, \cite{jia2012recognition} explored the dynamic updating of each cell in the clutter map based on changes in a specific environment. A series of related studies have enabled the clutter map to adapt to dynamic sensing environments. However, existing clutter map researches predominantly focus on the signal strength of two-dimensional (2D) ground clutter, with limited attention given to dynamic clutter and slow-moving targets in complex low-altitude environments \cite{gogineni2022high}. In addition, clutter map is mainly applied to mono-static radar sensing, and its extension to bi-static or multi-static collaborative sensing is challenging due to the difficulties of constructing, storing, and updating multiple delay-angle-intensity clutter maps in real time. 

Different from the standard-alone radar systems, the cooperation between BS and user equipment (UE) in ISAC systems provides new opportunities for clutter suppression. With the explosive growth in the number of BSs, UEs and cooperating nodes in future sixth-generation (6G) mobile communication networks, which greatly supports massive IoT connections, environment-aware clutter suppression is becoming more feasible. To this end, we propose a novel environment-aware clutter suppression technique based on the channel knowledge map (CKM) concept \cite{zeng2021toward}. As a site-specific database trying to learn the location-specific channel knowledge priors, CKM enables environment-awareness and is extremely valuable for wireless communication, localization and sensing in complex environment \cite{ckm_tutorial}. Unlike clutter maps that solely focus on clutter intensity, CKM offers a broader range of channel knowledge options, including but not limited to beam index, channel gain, direction of arrival (DoA), and the presence or absence of LoS paths. Existing CKM-based work includes model-based CKM construction \cite{xu2024much}\cite{qiu2024channel}, AI-based CKM construction \cite{11184538}\cite{dai2025generating}\cite{fu2025ckmdiff}, CKM-enabled beam alignment \cite{wu2023environment}\cite{taghavi2023environment}\cite{dai2024prototyping}, communication resource allocation \cite{zhan2023aerial}\cite{yue2024channel}, positioning and tracking \cite{wu2025you}\cite{10829585}.

In this paper, we propose a specialized CKM tailored for low-altitude UAV ISAC in complex IoT environment, termed clutter angle map (CLAM). CLAM provides location-specific prior angular information about primary clutter in the three-dimensional (3D) environment. A method for suppressing clutter based on DoA information provided by CLAM is proposed, which eliminates primary clutter in the spatial domain. To further account for scenarios when the sensing targets and clutters have close directions, where pure CLAM-enabled spatial domain method becomes less effective, we propose a two-step CLAM-enabled joint spatial–Doppler domain clutter suppression algorithm. The main contributions of this paper are listed as follows:
\begin{itemize}
    \item First, we present the detailed system model for low-altitude UAV ISAC in complex clutter environment. The traditional Doppler-based radar clutter suppression methods are then presented. The MTI techniques, which are extensively studied in the radar field, are then extended to the OFDM ISAC system. We then demonstrate that MTI techniques are effective in sensing high-speed targets, but face challenges for slow-moving targets.
    
    \item Second, a novel CLAM-enabled spatial domain clutter suppression method is proposed, which addresses the aforementioned limitations of Doppler-based approaches. Leveraging the clutter angle prior information provided by CLAM, spatial domain zero-forcing (ZF) is applied to suppress the clutter signals corresponding to clutter DoAs based on the UE location, while preserving desired signals scattered by the sensing targets. This operation enhances the signal-to-clutter ratio (SCR) and rejects clutter before subsequent signal processing and target sensing algorithms are applied.
    
    \item Third, to address close-angle scenarios where targets and clutter share similar DoAs and pure spatial domain clutter suppression becomes ineffective, a two-step CLAM-enabled joint spatial–Doppler domain clutter suppression algorithm is proposed. By sequentially validating the angle estimation results under spatial domain method and treating all other angles' signals as relative ‘clutter’, this approach distinguishes the desired targets in the joint delay-Doppler domain. Extracting signals of specific delay-Doppler pairs after angle ZF of ‘clutter’ enables more precise angle estimation and more comprehensive target sensing. This method avoids the influence of angle coupling at close angles and achieves clutter suppression in the joint domain.

    \item The effectiveness of our proposed algorithm was validated through extensive numerical simulations. A 3D low-altitude scenario was constructed, featuring static and dynamic clutter, fast-moving and slow-moving targets. Simulation results demonstrate that our proposed CLAM-enabled clutter suppression algorithm effectively suppresses clutter and enhance target sensing accuracy. Compared to traditional sensing methods without or with pure spatial/Doppler domain clutter suppression, our proposed joint domain approach achieves improved target sensing performance.
\end{itemize}

The rest of this paper is organized as follows. Sec.~\ref{System Model} introduces the system model of an uplink bi-static ISAC system in a low-altitude UAV environment. The classic Doppler-based clutter suppression methods and the preprocessing of signal to apply MTI to the OFDM waveform are analyzed in Sec.~\ref{Conventional MTI-based Clutter Suppression}. Then, Sec.~\ref{Clutter Suppression Enabled By CLAM} introduces CLAM-enabled joint spatial-Doppler domain clutter suppression algorithm. Simulation results are discussed in
Sec.~\ref{numerical simulation}. Finally, we conclude the paper in Sec.~\ref{Conclusion}.

\emph{Notation:} Scalars are denoted by italic letters. Vectors
and matrices are denoted by boldface lower- and uppercase
letters, respectively. $\mathbb{C}^{M \times N}$ represents the
space of $M \times N$ complex-valued matrices. $\boldsymbol{I}_N$ denotes an $N \times N$ identity matrix. The inverse, transpose, Hermitian, conjugate and determinant are respectively denoted by $(\cdot)^{-1}$, $(\cdot)^{T}$, $(\cdot)^{H}$, $(\cdot)^{*}$ and $|\cdot|$. The Frobenious-norm of the matrix $\boldsymbol{A}$ is given by $||\boldsymbol{A}||_F$. The Kronecker product and the linear convolution are denoted by $\otimes$ and $\ast$, respectively. The modulo operation with respect to $X$ is represented by $\mathrm{mod}_X(\cdot)$, where $X$ is a positive integer. $\delta(\cdot)$ denotes the Dirac delta function. The operator $\lfloor\cdot\rfloor$ denotes the integer floor operation. For two sets $\mathcal{A}$ and $\mathcal{B}$, $\mathcal{A} \backslash \mathcal{B}$ denotes their set difference.

\section{System Model}
\label{System Model}
\begin{figure}[tbp]
    \centering    
    \includegraphics[width=0.95\columnwidth]{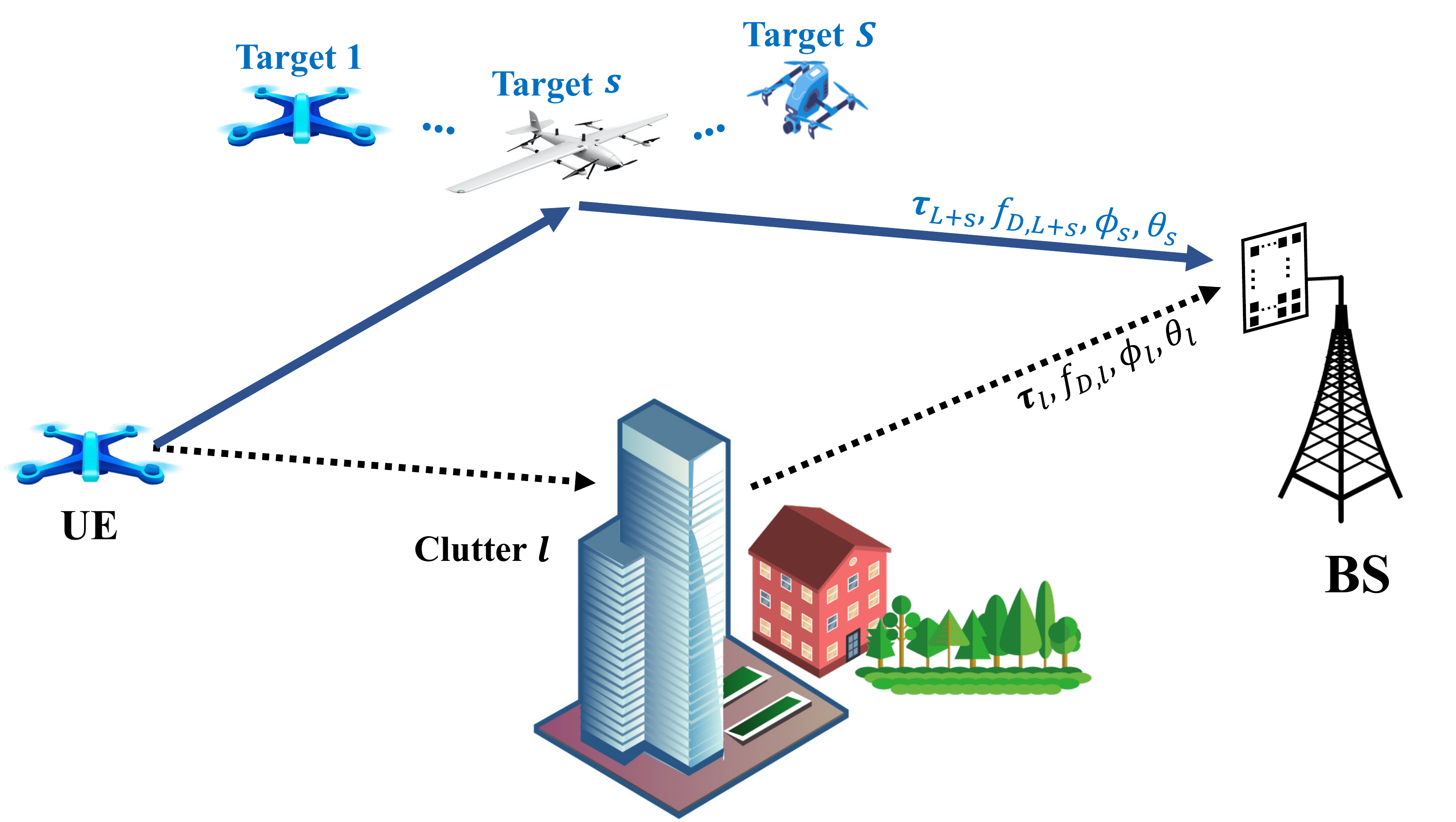}
    \caption{An uplink bi-static ISAC system in complex low-altitude environment.}
    \label{scenary}
\end{figure}

As shown in Fig.\;\ref{scenary}, we consider a single-input multiple-output (SIMO) bi-static ISAC system. To support UE communication and UAV sensing in a 3D low-altitude environment, the BS is equipped with a uniform planar array (UPA) of $M = M_x \times M_z$ antennas and tries to sense $S$ UAVs simultaneously. Assume that there are $L$ major clutters in the low-altitude environment. We can decompose the channel into the channel by the background  clutter and that by the targets of interest:
\begin{equation}
\begin{aligned}
    \label{eq1}
    \boldsymbol{h}(t,\tau )=&\sum_{l=1}^{L}{\boldsymbol{h}_l\delta \left( \tau -\tau_l \right) e^{j2\pi f_{D,l}t}}\\
    &+\sum_{s=L+1}^{L+S}\boldsymbol{h}_s\delta \left( \tau -\tau_s \right) e^{j2\pi f_{D,s}t},
\end{aligned}
\end{equation}
where $t$ and $\tau$ denote the time and delay, respectively. The channel coefficient vector of the $l^{th}$ clutter path is denoted as $\boldsymbol{h}_l=\beta_l\boldsymbol{\alpha}(\phi_l, \theta_l)$, $l=1, \cdots, L$. $\boldsymbol{h}_s=\beta_s\boldsymbol{\alpha}(\phi_s, \theta_s)$ is the channel coefficient vector of the $s^{th}$ path from the target, where $s=L+1,\cdots, L+S$. $\boldsymbol{\alpha}( \phi, \theta)\in \mathbb{C}^{M \times 1}$ represents the array response vector of a signal arriving with an azimuth angle $\phi$ and a zenith angle $\theta$. $\{\beta_l,\beta_s\}$ and $\{\tau_l,\tau_s\}$ are the corresponding path gains and propagation delays, respectively. $f_{D,k}, k=1,\cdots, L+S$ are the Doppler frequency caused by the scatterers and the moving UAV targets.

For a signal with an azimuth angle $\phi$ and a zenith angle $\theta$, the UPA's array response vector can be expressed as the Kronecker product of steering vectors of two dimensions:
\begin{equation}
    \boldsymbol{\alpha } (\phi, \theta) = \boldsymbol{\alpha }_x(\phi, \theta) \otimes \boldsymbol{\alpha }_z(\theta) \in \mathbb{C}^{M \times 1},
\end{equation}
where $\boldsymbol{\alpha }_x(\phi, \theta)$ and $\boldsymbol{\alpha }_z(\theta)$ are
\begin{align}
    \label{a_x}
    \boldsymbol{\alpha }_x(\phi, \theta) &= 
    \left[ e^{j 2 \pi m_x \frac{d}{\lambda}\cos \phi \sin \theta} \right]_{m_x=0}^{M_x-1} \in \mathbb{C}^{M_x \times 1}, \\   
    \label{a_z}
    \boldsymbol{\alpha }_z(\theta) &= \left[ e^{j 2 \pi m_z \frac{d}{\lambda}\cos \theta} \right]_{m_z=0}^{M_z-1} \in \mathbb{C}^{M_z \times 1}.
\end{align}

In \eqref{a_x} and \eqref{a_z}, $\lambda=c/f_c$ is the wavelength of the signal and $d$ represents the antenna spacing. $c$ denotes the speed of light while $f_c$ represents the carrier frequency. 

The received signal at the BS is
\begin{equation}
    \label{eq2}
    \boldsymbol{y}\left( t \right) =\int_{0}^{\infty}{s\left( t-\tau \right) \boldsymbol{h}\left( t,\tau \right) d\tau}+\boldsymbol{n}\left( t \right),
\end{equation}
where $s(t)$ is the transmitted signal from the UE and $\boldsymbol{n}(t)\in \mathbb{C}^{M\times 1}$ is received additive Gaussian noise. Substituting \eqref{eq1} into \eqref{eq2}, we have
\begin{equation}
\label{y(t)}
\begin{aligned}
    \boldsymbol{y}(t)=&\sum_{l=1}^{L}{\boldsymbol{h}_ls\left( t-\tau _l \right) e^{j2\pi f_{D,l}t}}\\
    &+\sum_{s=L+1}^{L+S}\boldsymbol{h}_ss\left( t-\tau _s \right) e^{j2\pi f_{D,s}t}+\boldsymbol{n}(t),
\end{aligned}
\end{equation}
where the first term is the clutter signal at the BS to be suppressed, and the second term is the useful signal reflected by the sensing targets. The BS aims to estimate the DoA $\{(\phi_s, \theta_s)\}_{s=L+1}^{L+S}$, delay $\{\tau_s\}_{s=L+1}^{L+S}$, and Doppler $\{f_{D,s}\}_{s=L+1}^{L+S}$ of the targets from the second term in \eqref{y(t)}. Therefore, it is necessary to enhance target sensing by distinguishing between clutter and target signals through clutter suppression.


\section{MTI-based Clutter Suppression}
\label{Conventional MTI-based Clutter Suppression}

If all clutter components are static with zero Doppler, i.e., $f_{D,l}=0$, complete static clutter suppression can be achieved by directly performing delay-Doppler estimation on the signal and setting the column at $f_{D}=0$ to zero\cite{110CWhitePaper}. However, such simple ZF cannot suppress the side lobes induced by the clutter in the Doppler domain. Furthermore, due to UE movement and the perturbation of scatterers, it is also possible that the clutters may have non-zero Doppler, i.e., $|f_{D,l}|>0$ may occur. Therefore, to better suppress the side lobes of zero-Doppler clutters and those with low-Doppler, MTI methods are often employed for more refined clutter suppression.

\subsection{MTI for Classic Periodic Radar Signals}
\label{MTI-perodic}

For calssic periodic radar signals, the Doppler-based clutter suppression methods usually assume that clutter signals are quasi-static with $f_{D,l}=0$ or $|f_{D,l}| \ll |f_{D,s}|$. Therefore, clutter suppression can be performed by high-pass filtering in the Doppler domain. According to \cite{mti, MTI_paper2}, in the time domain, the impulse response of  an $N$th-order MTI filter can be represented as a series of weighted impulses:
\begin{equation}
\label{h_MTI}
    h_{\mathrm{MTI}}(t)=\sum_{n=0}^{N} c_{n} \delta(t-n T),
\end{equation}
where $T$ is the sampling interval. In a radar system, $T$ is generally set as the pulse repetition interval (PRI), or adjusted according to engineering requirements \cite{T_designing}. $\{c_n\}_{n=0}^N$ represent the filter coefficients to be designed. Then, the MTI filter's frequency response can be expressed as
\begin{equation}
\label{H_MTI}
    H_{\mathrm{MTI}}(f)=\mathcal{F}\left\{h_{\mathrm{MTI}}(t)\right\}=\sum_{n=0}^{N} c_{n} e^{-j 2 \pi f n T}.
\end{equation}

To suppress the quasi-static clutter signals, the MTI filter is designed to be a high-pass filter, with zero response at DC, i.e., $H_{\mathrm{MTI}}(0)=0$, which is achieved by choosing the coefficients such that $\sum_{n=0}^{N}c_n=0$. Practical designs of coefficients $\{c_n\}_{n=0}^N$ should depend on filter type, order, cut-off frequency, and other design constraints \cite{ Adaptive-MTI-design, Optimum-MTI-design-1, Optimum-MTI-design-2}.

The received signal passing through the MTI high-pass filter can be written as the convolution of the filter's impulse response with the time domain signal:
\begin{equation}
\label{y_MTI}
    \boldsymbol{y}_{\mathrm{MTI}}(t)=\boldsymbol{y}(t) \ast h_{\mathrm{MTI}}(t)=\sum_{n=0}^{N} c_{n} \boldsymbol{y}(t-n T).
\end{equation}

According to \eqref{y(t)}, the frequency domain representation of the signal is 
\begin{equation}
\begin{aligned}
\label{y(f)}
\boldsymbol{y}^{\mathrm{rev}}(f) & =\sum_{l=1}^{L} \boldsymbol{h}_{l} s\left(f-f_{D, l}\right) e^{-j 2 \pi\left(f-f_{D, l}\right) \tau_{l}} \\
& +\sum_{s=L+1}^{L+S} \boldsymbol{h}_{s} s\left(f-f_{D, s}\right) e^{-j 2 \pi\left(f-f_{D, s}\right) \tau_{s}}+ \boldsymbol{n}^{\mathrm{rev}}(f),
\end{aligned}
\end{equation}
where $s(f)$ is the frequency domain representation of the signal transmitted by the UE. $\boldsymbol{n}^{\mathrm{rev}}(f)$ denotes the spectrum of the received noise.

With \eqref{H_MTI} and \eqref{y(f)}, we obtain the frequency domain representation of the MTI output:
\begin{equation}
\begin{aligned}
\label{y_MTI^out}
\boldsymbol{y}_{\mathrm{MTI}}^{\mathrm{out}}&(f)=H_{\mathrm{MTI}}(f) \boldsymbol{y}^{\mathrm{rev}}(f)\\
&=\sum_{l=1}^{L} \boldsymbol{h}_{l} s\left(f-f_{D, l}\right) e^{-j 2 \pi\left(f-f_{D, l}\right) \tau_{l}} H_{\mathrm{MTI}}(f) \\
& +\sum_{s=L+1}^{L+S} \boldsymbol{h}_{s} s\left(f-f_{D, s}\right) e^{-j 2 \pi\left(f-f_{D, s}\right) \tau_{s}} H_{\mathrm{MTI}}(f) \\
& + \boldsymbol{n}_{\mathrm{MTI}}^{\mathrm{out}}(f),
\end{aligned}
\end{equation}
where $\boldsymbol{n}_{\mathrm{MTI}}^{\mathrm{out}}(f) = \boldsymbol{n}^{\mathrm{rev}}(f)H_{\mathrm{MTI}}(f)$ is the filtered noise. 

Equation \eqref{y_MTI^out} indicates that for each frequency component of the signal, there is a corresponding response of the MTI filter, which either retains or suppresses the signal. Therefore, we focus on the characteristics of the frequency response of the MTI filter in \eqref{H_MTI}: $(1)$ \emph{Periodicity}. The MTI filter response is periodic in $f$ with period $1/T$, i.e., $H_{\mathrm{MTI}}(f+m/T)=H_{\mathrm{MTI}}(f), m\in \mathbb{Z}$. $\sum_{n=0}^{N} c_{n}=0$ ensures complete suppression of the zero frequency, so \eqref{H_MTI} describes a comb filter whose frequency response has zeros at integer multiples of $1/T$. $(2)$ \emph{High-pass filter characteristic}. Within each period, the MTI filter exhibits high-pass filtering characteristics, suppressing low frequencies and retaining high frequencies. The spectrum of periodic classic radar signals consists of discrete spectral lines that are equally spaced at intervals of $1/T$ and lie exactly at the periodic nulls of the corresponding high-pass filter. 

\begin{figure}[tbp]
    \centering
    \includegraphics[width=0.95\linewidth]{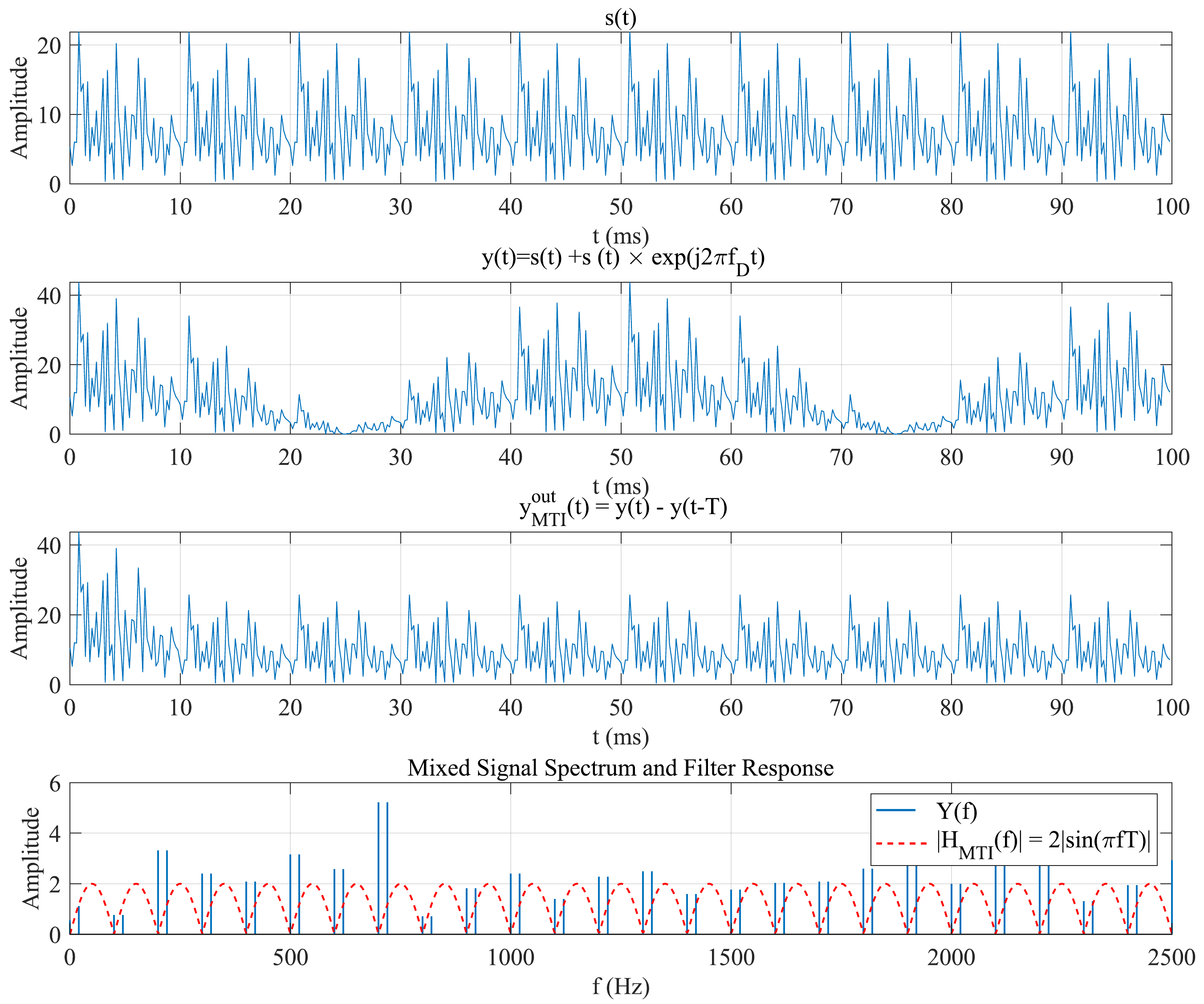}
    \caption{The time domain and frequency domain responses of a first-order MTI filter for clutter suppression on periodic radar signals.}
    \label{4pics}
\end{figure}
Taking the classic periodic pulsed signal of radar in Fig.\;\ref{4pics} as an example, $s(t)$ is a periodic signal with a period $T = 10\,\mathrm{ms}$, regarded as the static clutter signal. After being scattered by moving targets, the Doppler frequency $f_D = 20$Hz distorts the received signal $y(t)=s(t)+s(t)e^{j2\pi f_Dt}$ compared to the clutter $s(t)$, where the noise is ignored for simplicity. With the simplest first-order filter ($c_0=1$, $c_1=-1$), the output time domain signal $y_{\mathrm{MTI}}^{\mathrm{out}}(t) = y(t) - y(t-T)$ effectively removes the zero-Doppler clutter $s(t)$. However, a Doppler frequency will slightly shift the clutter $s(t)$'s spectral lines away from the zeros of the filter's frequency response. The amount of this offset determines the extent to which the target signal is preserved. The specific degree of retention is determined by the filter's response, namely $\{c_n\}_{n=0}^N$ and the order $N$.

Specifically, as shown in Fig.\;\ref{4pics}, $\mathbf{c}=[1, -1]$ creates the simplest MTI filter. Common airborne or ground radar clutter suppression employs such first- to third-order MTI filters. A more sophisticated filter design would consider the received signals within multiple time periods and perform complex, weighted phase cancellation on them. However, designing a filter that meets practical needs in different scenarios is challenging, especially in the absence of environmental information. This will be further analyzed in Sec.~\ref{Limitations}.



Note that the design of MTI has exploited the periodicity of the radar signal, which is, however, not true for the continuous waveform, such as OFDM used in ISAC. Moreover, directly applying filter convolution to OFDM can introduce severe inter-symbol interference (ISI). Consequently, additional signal processing is required when applying MTI to OFDM signals, as detailed in the next subsection.

\subsection{MTI for OFDM Signals}
\label{MTI+OFDM}

In this subsection, we propose a new method for MTI clutter suppression in the symbol domain for non-periodic OFDM waveforms. The core idea is to perform MTI high-pass filtering in Sec.~\ref{MTI-perodic} on the reconstructed OFDM signal in the symbol domain. 
Following \cite{dai2025tutorial}, the time domain transmitted OFDM signal can be written as:
\begin{equation}
\label{OFDM_s(t)}
    \begin{aligned}
        s(t) =&\sum_{\gamma = 0}^{N_{\mathrm{sym}} - 1} \sum_{n=0}^{N_{\mathrm{sc}}-1} b_{n,\gamma} e^{j2\pi n\Delta f(t-\gamma T_{s}-T_{\mathrm{CP}})}\\
        &\times \mathrm{rect} \left(\frac{t-\gamma T_s}{T_s}-\frac{1}{2}\right),
    \end{aligned}
\end{equation}
where $N_{\mathrm{sym}}$ and $ N_{\mathrm{sc}}$ denote the number of transmitted symbols and subcarriers, respectively. $\Delta f$ is the subcarrier spacing. $b_{n,\gamma}$ is the data symbol on the $n^{th}$ subcarrier and the $\gamma^{th}$ OFDM symbol. $T_{\mathrm{CP}}$ and $T_{\mathrm{sym}}$ represent the duration of the cyclic prefix (CP) and the OFDM symbol without CP. Thus the OFDM symbol duration including CP is $T_s= T_{\mathrm{CP}} + T_{\mathrm{sym}}$. $\mathrm{rect}(t)$ denotes the unit rectangular window, equal to 1 for $t\in\left[-\tfrac12,\tfrac12\right]$ and 0 otherwise.

Substituting \eqref{OFDM_s(t)} into \eqref{y(t)}, the received OFDM sensing signal $\boldsymbol{y}_{\mathrm{RX}}(t)\in \mathbb{C}^{M\times 1}$ at the BS is
\begin{small}
\begin{equation}
\begin{aligned}
\label{y_s(t)}
    &\boldsymbol{y}_{\mathrm{RX}}(t)= \\
    &\sum_{l=1}^{L}\sum_{\gamma = 0}^{N_{\mathrm{sym}} - 1} \sum_{n=0}^{N_{\mathrm{sc}}-1} \boldsymbol{h}_lb_{n,\gamma} e^{j2\pi n\Delta f(t-\tau_l-\gamma T_{s}-T_{\mathrm{CP}})}e^{j2\pi f_{D,l}t}\Pi_{l,\gamma}(t) + \\   
    &\sum_{s=L+1}^{L+S}\sum_{\gamma = 0}^{N_{\mathrm{sym}} - 1} \sum_{n=0}^{N_{\mathrm{sc}}-1} \boldsymbol{h}_sb_{n,\gamma} e^{j2\pi n\Delta f(t-\tau_s-\gamma T_{s}-T_{\mathrm{CP}})}e^{j2\pi f_{D,s}t}\Pi_{s,\gamma}(t) \\
    &+\boldsymbol{n}(t).
\end{aligned}
\end{equation}
\end{small}

In \eqref{y_s(t)}, $\Pi_{k,\gamma}(t)=\mathrm{rect} \left(\frac{t-\tau_k-\gamma T_s}{T_s}-\frac{1}{2}\right)$ is the rectangular window for the  $k^{th}$ path, at the $\gamma^{th}$ symbol index with respect to time $t$, which equals $1$ only during the interval $[\tau_k + \gamma T_s, \tau_k + (\gamma+1) T_s]$. Then, we can divide \eqref{y_s(t)} into $N_{\mathrm{sym}}$ blocks, each of a time duration of $T_s$. After CP removal, for $\gamma=0,\cdots, N_{\mathrm{sym}}-1$, the $\gamma^{th}$ block can be represented as
\begin{equation}
    \boldsymbol{y}^{\gamma}_{\mathrm{RX}}(t)=\boldsymbol{y}_{\mathrm{RX}}(t+\gamma T_s+T_{\mathrm{CP}}) \mathrm{rect}\left(\frac{t}{T_{\mathrm{sym}}}-\frac{1}{2}\right).
\end{equation}

The condition $\mathrm{max}\{\tau_l, \tau_s\}\leq T_{\mathrm{CP}}$ is typically enforced at design time to avoid ISI. Under the assumption that $\Delta f \gg |f_{D,k}|, k=1, \cdots, L+S$, the inter-carrier interference (ICI) caused by a large Doppler frequency shift can be neglected and the Doppler-induced phase shift remains constant within a single OFDM symbol \cite{dai2025tutorial}. After $N_{\mathrm{sc}}$-point sampling of $\boldsymbol{y}_{\mathrm{RX}}^{\gamma}(t)$, the $q^{th}$ sample of the $\gamma^{th}$ block is $\boldsymbol{y}^{\gamma}_{\mathrm{RX}}[q]=\boldsymbol{y}^{\gamma}_{\mathrm{RX}}(q/B)$, $q=0,\cdots, N_{\mathrm{sc}}-1$, where $B=N_{\mathrm{sc}}\Delta f$ denotes the sampling rate. Then, for signal processing, we can perform an $N_{\mathrm{sc}}$-point DFT on each block $\boldsymbol{y}^{\gamma}_{\mathrm{RX}}[n]$ according to \cite{dai2025tutorial}, obtaining the frequency domain signal for the $n^{th} \left(n=0,\cdots, N_{\mathrm{sc}}-1\right)$ sample of the $\gamma^{th} \left(\gamma=0,\cdots, N_{\mathrm{sym}}-1\right)$ OFDM block:
\begin{equation}
\begin{aligned}
\label{y_RX^gamma[n]}
    \bar{\boldsymbol{y}}_{\mathrm{RX}}^{\gamma}[n]&=\frac{1}{N_{\mathrm{sc}}} \sum_{q=0}^{N_{\mathrm{sc}}-1} \boldsymbol{y}_{\mathrm{RX}}^{\gamma}[q] e^{-j 2 \pi n q / N_{\mathrm{sc}}}\\
    &\approx b_{n, \gamma}\left(\sum_{l=1}^{L} \tilde{\boldsymbol{h}}_{l} e^{-j 2 \pi n \Delta f \tau_{l}} e^{j 2 \pi \gamma T_{s} f_{D, l}} \right. \\
    &\left. +\sum_{s=L+1}^{L+S} \tilde{\boldsymbol{h}}_{s} e^{-j 2 \pi n \Delta f \tau_{s}} e^{j 2 \pi \gamma T_{s} f_{D, s}}\right) + \bar{\boldsymbol{n}}^{\gamma}[n],
\end{aligned}
\end{equation}
where $\tilde{\boldsymbol{h}}_{l} \triangleq \boldsymbol{h}_{l} e^{j 2 \pi f_{D,l} T_{\mathrm{CP}}}$ and $\tilde{\boldsymbol{h}}_{s} \triangleq \boldsymbol{h}_{s} e^{j 2 \pi f_{D,s} T_{\mathrm{CP}}}$. $\bar{\boldsymbol{n}}^{\gamma}[n]$ represents the corresponding noise in the frequency domain. Noting that in \eqref{y_RX^gamma[n]}, since $\Delta f \gg |f_{D,k}|$, $e^{j 2 \pi f_{D,k}q/B} \approx 1$ can be approximated for $|f_{D,k}q/B|<|f_{D,k}N_{sc}/B|=|f_{D,k}|/\Delta f \ll 1$. Consequently, by combining these sampled values, we can form an $ M \times N_{\mathrm{sc}} \times N_{\mathrm{sym}} $ dimensional tensor to reflect the signal in the spatial-frequency-symbol domain. The tensor $\boldsymbol{Y}_{\mathrm{RX}} \in \mathbb{C}^{M \times N_{\mathrm{sc}} \times N_{\mathrm{sym}}} $ can be defined by
\begin{equation}
    \boldsymbol{Y}_{\mathrm{RX}}(:, n, \gamma) = \bar{\boldsymbol{y}}_{\mathrm{RX}}^{\gamma}[n].
\end{equation}

Since the random data symbols $b_{n,\gamma}$ are related to the frequency and symbol domains, they will affect the manifold structure and hinder the complete suppression of zero-Doppler signals. As a result, we need to remove $b_{n,\gamma}$ from \eqref{y_RX^gamma[n]} before further processing, which can be assumed to has been correctly decoded. Therefore, by element-wise division of each data symbol from $\boldsymbol{Y}_{\mathrm{RX}}$, we obtain the following tensor:
\begin{equation}
\begin{aligned}
\label{Y-3D}
    \boldsymbol{Y}_{\mathrm{div}}(:, n, \gamma) &= \frac{\boldsymbol{Y}_{\mathrm{RX}}(:, n, \gamma)}{b_{n,\gamma}}\\
    & = \sum_{l=1}^{L} \tilde{\boldsymbol{h}}_{l} e^{-j 2 \pi n \Delta f \tau_{l}} e^{j 2 \pi \gamma T_{s} f_{D, l}}\\
    &+\sum_{s=L+1}^{L+S} \tilde{\boldsymbol{h}}_{s} e^{-j 2 \pi n \Delta f \tau_{s}} e^{j 2 \pi \gamma T_{s} f_{D, s}}+\tilde{\boldsymbol{n}}^{\gamma}[n], 
\end{aligned}    
\end{equation}
where $\tilde{\boldsymbol{n}}^{\gamma}[n] = \bar{\boldsymbol{n}}^{\gamma}[n] / b_{n,\gamma}$ denotes the resulting noise.

Based on this, a digital filter of order $N$ can be designed and applied to the symbol domain of $\boldsymbol{Y}_{\mathrm{div}}$, by
\begin{equation}
\label{OFDM_MTI}
    \bar{\boldsymbol{Y}}(:, :, \gamma) = \sum_{k=0}^{N} c_{k}\boldsymbol{Y}_{\mathrm{div}}(:, :, \gamma-k) \in \mathbb{C}^{M \times N_{\mathrm{sc}} \times N_{\mathrm{sym}}},
\end{equation}
where we set $\boldsymbol{Y}_{\mathrm{div}}(:, :, \gamma)=0$, $(\gamma=-N,-N+1,\cdots,-1)$  to ensure proper boundary handling. 

Similarly, the filter coefficients applied across OFDM symbols must also satisfy $\sum_{n=0}^{N} c_{n}=0$ to ensure complete suppression of zero-Doppler clutter. \eqref{y_MTI} and \eqref{OFDM_MTI} are intrinsically equivalent, as both perform the convolution of signals separated by integer multiples of a period in the time domain. The distinction lies in their application: \eqref{y_MTI} targets the classic periodic radar signals in the time domain, whereas \eqref{OFDM_MTI} applies convolution to consecutive OFDM symbols after demodulation and element-wise division.

\subsection{Limitations of Doppler-based techniques}
\label{Limitations}
\begin{figure}[tbp]
    \centering
    \includegraphics[width=0.9\linewidth]{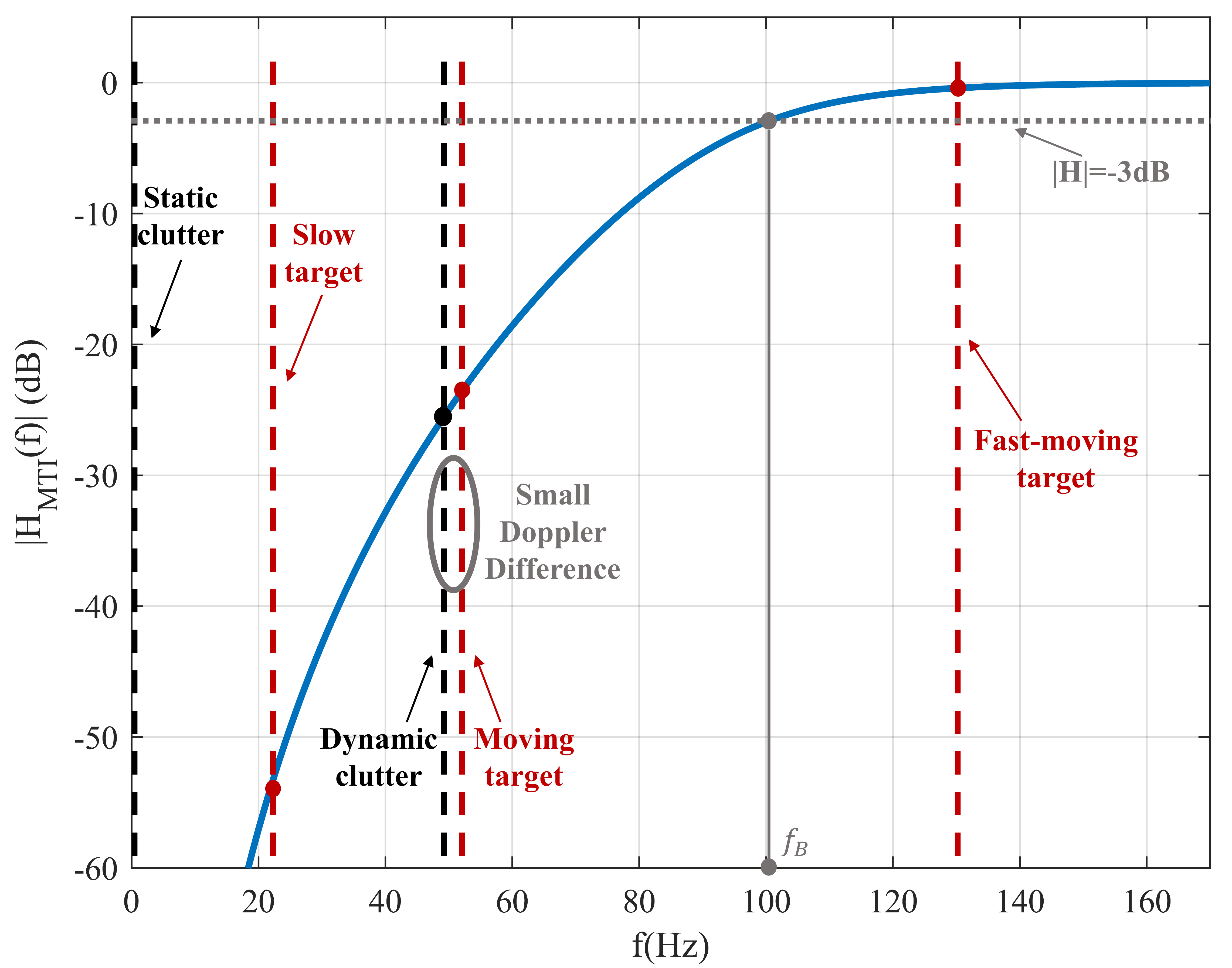}
    \caption{The frequency response of a Doppler high-pass filter and the Doppler distributions of different clutter and targets.}
    \label{HPF}
\end{figure}

From the frequency domain perspective, a Doppler shift $f_D$ offsets the spectrum from its original position. Therefore, rather than considering the entire response $|H_{\mathrm{MTI}}(f)|$ in \eqref{y_MTI^out}, it suffices to examine the sampled magnitude $|H_{\mathrm{MTI}}(f_D)|$ to assess how an MTI filter affects a signal with Doppler $f_D$. Define $f_B$ as the 3\,dB cut-off frequency of the MTI filter that satisfies $\left|H_{\mathrm{MTI}}\left(f_{B}\right)\right|=1 / \sqrt{2}$ and $0<|f_B|<\frac{1}{2T}$. We consider $f_B$ to be a critical Doppler value that determines whether the signal is suppressed or retained. The specific value of $f_B$ depends on the technical specifications of the actual design, which is ultimately reflected in $\{c_n\}_{n=0}^N$. Fig.\;\ref{HPF} shows the frequency amplitude response of a Doppler high-pass filter and the Doppler distributions of different clutter and targets. Then, the impact of $|H_{\mathrm{MTI}}(f_D)|$ for different $f_D$ is summarized as follows:
\begin{itemize}
    \item $f_D\rightarrow0$. For static clutter whose $f_D\rightarrow0$, the corresponding $|H_{\mathrm{MTI}}(f_D)|\rightarrow0$, ensuring the suppression of zero-Doppler clutter.
    

    \item $0<|f_D|<|f_B|$. For clutter or target with a relatively small Doppler, the spectrum lies within the transition band of the high-pass filter. Therefore, clutter suppression methods based on Doppler selection suppress both clutter and targets with small Doppler frequencies. 

    \item $|f_D| \gg |f_B|$. Within the passband of the filter, it can be approximated that $|H_{ \mathrm{MTI} }(f_D)| \approx 0$\,dB. Thus, fast-moving targets are well preserved.
\end{itemize}

In summary, Doppler-based high-pass filtering methods share the following limitations. 

\begin{itemize}
    \item \textbf{\emph{Insufficient Doppler resolution}}. When the Doppler difference between a target and dynamic clutter is very small, high Doppler resolution is required for the filter to distinguish them. However, filters often cannot meet such requirements in practice. If the coherence processing interval (CPI) is too short, the Doppler resolution will be inadequate to distinguish targets from clutter. If it is excessively long, computational load and real-time responsiveness become problematic.

    \item \textbf{\emph{Difficulty of filter design}}. Practical filter design must account for the clutter Doppler range, cutoff frequency, and transition bandwidth. Without prior clutter information, these quantities are hard to obtain. This poses significant challenges for filter design. Moreover, even with prior information on the Doppler distribution of clutter, it is difficult to strike a balance between suppressing dynamic clutter and retaining slow-moving low-altitude targets. This is also verified in the simulation results presented in Sec.~\ref{MTI-sim}.
\end{itemize}

Therefore, more effective clutter-suppression schemes in other domains are needed. Since major clutter components in a bi-static environment have rather stable DoAs, suppression in the spatial domain is a natural option. Based on this, in the following, we propose a CLAM-enabled spatial domain clutter suppression, as well as its extension to joint spatial–Doppler domain suppression.

\section{Environment-Aware Clutter Suppression Enabled by CLAM}
\label{Clutter Suppression Enabled By CLAM}
If the DoAs of the major clutters are known in advance, the BS can perform spatial ZF to suppress these clutter DoAs and hence effectively eliminate the influence of clutter from these directions. CKM offers location-specific channel knowledge based on the transceiver positions, which is exploited as a priori information for subsequent processing. In the following, one instance of CKM, namely CLAM, is introduced to provide the clutter angle information for clutter suppression in low-altitude OFDM ISAC.

\subsection{Basic Principles of \rm{CLAM}}
CLAM provides the DoAs of background clutter for potential UE locations by offline measurements of BSs. Specifically, let $\mathbf{P}$ denote the location space containing all potential UE positions, and let a particular location be $\mathbf{p}_{\mathrm{UE}}$. The CLAM, denoted by $\mathcal{M}_{\mathrm{clutter}}$, is a mapping from location to clutter angles:
\begin{equation}
    \label{M_clutter}
    \mathcal{M}_{\mathrm{clutter}}: \mathbf{p}_{\mathrm{UE}}\in \mathbf{P}\rightarrow \Theta \left( \mathbf{p}_{\mathrm{UE}} \right) ,
\end{equation}
where $\Theta \left( \mathbf{p}_{\mathrm{UE}} \right)$ is the corresponding clutter angle information, containing $L$ clutter DoAs as 
\begin{equation}
    \label{Theta}
    \Theta \left( \mathbf{p}_{\mathrm{UE}} \right) =\{ (\phi _l,\, \theta_l)  \}_{l=1}^{L},
\end{equation}
where $\phi_l$ and $\theta_l$ are the azimuth and zenith angle of the $l^{th}$ clutter path. 

For CLAM construction, a straightforward approach is to sample all UE locations in the area of interest and, at each location, estimate the $L$ clutter DoAs using standard angle-estimation algorithms such as periodogram, multiple signal classification (MUSIC) and orthogonal matching pursuit (OMP) \cite{dai2025tutorial}. Since the possible UE locations are continuous, a more practical method is to divide the space into grids or irregular cells, with each small area corresponding to a specific clutter angle \cite{dai2024prototyping}. To further complete the map from a limited number of measured locations, we can interpolate across cells or leverage learning-based methods (e.g., deep or generative models) to predict clutter-angle priors at arbitrary UE positions \cite{11184538,fu2025ckmdiff}. In this way, with the prior clutter information offered by CLAM, the system is able to achieve clutter environment-awareness. In the next section, we will specify how the clutter angle information can be utilized to achieve spatial domain clutter suppression. 

\subsection{CLAM-enabled Spatial Domain Clutter Suppression}
\label{CLAM-based Spatial Clutter Suppression}

For DoA estimation in OFDM-ISAC, we can treat the sensing data in the subcarrier and symbol domains as snapshots \cite{dai2025tutorial}. Then, the tensor $\boldsymbol{Y}_{\mathrm{div}}\in \mathbb{C}^{M \times N_{\mathrm{sc}} \times N_{\mathrm{sym}}}$ in \eqref{Y-3D} can be rewritten to an $M \times P$ matrix $\boldsymbol{Y}$, where $P = N_{\mathrm{sc}}N_{\mathrm{sym}}$ is the number of snapshot. The matrix $\boldsymbol{Y}\in \mathbb{C}^{M \times P}$ can be represented by
\begin{equation}
    \boldsymbol{Y}(:, p_{n,\gamma}) = \boldsymbol{Y}_{\mathrm{div}}(:, n, \gamma), \,\, p_{n, \gamma} = \gamma N_{\mathrm{sc}} + n,
\end{equation}
where $n \in [ 0, N_{\mathrm{sc}-1} ]$, $\gamma \in [ 0, N_{\mathrm{sym}-1} ]$ and $n,\gamma\in \mathbb{Z}$. Then, we can express $\boldsymbol{Y}$ as
\begin{equation}
\label{Y=AS+N}
    \boldsymbol{Y} = \boldsymbol{A}\boldsymbol{S} + \boldsymbol{N},
\end{equation}
where $\boldsymbol{N}\in \mathbb{C}^{M \times P}$ is the noise matrix. The manifold matrix with all received DoAs $\boldsymbol{A}\in \mathbb{C}^{M\times (L+S)}$ and the signal matrix  $\boldsymbol{S}\in \mathbb{C} ^{\left( L+S \right) \times P}$ can respectively be represented as
\begin{align}
    \boldsymbol{A} &= \left[\, \boldsymbol{\alpha }( \phi_1, \theta_1 ), \cdots, \boldsymbol{\alpha }( \phi_{L+S}, \theta_{L+S} ) \, \right] ,\\
    \boldsymbol{S}(k, p_{n, \gamma}) &= \beta_ke^{-j2 \pi n \Delta f \tau_k} e^{2 \pi f_{D,k}(\gamma T_s+T_{\mathrm{CP}})}.
\end{align}

The signal matrix $\boldsymbol{S}$ can be further decomposed into two parts:
\begin{equation}
\boldsymbol{S}=\left[ \begin{array}{c}
	\boldsymbol{S}_{\mathrm{c}}\\
        \boldsymbol{S}_{\mathrm{s}}\\
\end{array} \right],  
\end{equation}
where $\boldsymbol{S}_{\mathrm{c}}\in \mathbb{C}^{L\times P}$ consists of $L$ clutter signals, and $\boldsymbol{S}_{\mathrm{s}}\in \mathbb{C}^{S\times P}$ denotes the sensing signals from $S$ UAV targets. 

Assume that the UE position $\mathbf{p}_{\text{UE}}$ is known, we can retrieve the clutter direction information by querying the proposed CLAM in \eqref{M_clutter}, and then suppress the clutter signals from these directions in the spatial domain. Suppressing clutter signals not only enhances the target sensing performance but also significantly reduces the computational load. With the CLAM-enabled clutter angles in  \eqref{Theta}, the manifold matrix of UPA over the given clutter angles can be written as
\begin{equation}
    \boldsymbol{C} = \left[\, \boldsymbol{\alpha }( \phi_1, \theta_1), \boldsymbol{\alpha }( \phi_2, \theta_2 ),\cdots, \boldsymbol{\alpha }( \phi_L, \theta_L ) \, \right] \in \mathbb{C}^{M\times L}.
\end{equation}

The orthogonal projector onto the clutter subspace $\mathcal{R}(\boldsymbol{C})$ is $\boldsymbol{C} (\boldsymbol{C}^H\boldsymbol{C} ) ^{-1}\boldsymbol{C}^H\in \mathbb{C}^{M\times M}
$. Define the complementary (zero-forcing) projector $\boldsymbol{W}=\boldsymbol{I}_M-\boldsymbol{C}( \boldsymbol{C}^H\boldsymbol{C} ) ^{-1}\boldsymbol{C}^H\in \mathbb{C}^{M\times M}$. Applying the normalized  ZF matrix $\bar{\boldsymbol{W}}=\boldsymbol{W}/||\boldsymbol{W}||$ to $\boldsymbol{A}$ projects the signal onto the orthogonal complement of the clutter subspace $\mathcal{R}(\boldsymbol{C})$. Rewrite $\boldsymbol{A}$ as $\left[ \boldsymbol{C}, \boldsymbol{A}_s \right]$, where $\boldsymbol{A}_s\in \mathbb{C}^{M\times S}$ is the manifold matrix of UPA in targets' DoAs. Applying $\bar{\boldsymbol{W}}$ to the array output signal yields:
\begin{equation}
\begin{aligned}
\boldsymbol{Y}_{\mathrm{ZF}}&=\bar{\boldsymbol{W}}\boldsymbol{Y}
\\
&=\bar{\boldsymbol{W}}\left[ \boldsymbol{C}, \boldsymbol{A}_s \right] \left[ \begin{array}{c}
	\boldsymbol{S}_c\\
	\boldsymbol{S}_s\\
\end{array} \right] +\bar{\boldsymbol{W}}\boldsymbol{N}
\\
&=\left[ \boldsymbol{O}_{M \times L}, \bar{\boldsymbol{W}}\boldsymbol{A}_s \right] \left[ \begin{array}{c}
	\boldsymbol{S}_c\\
	\boldsymbol{S}_s\\
\end{array} \right] +\bar{\boldsymbol{W}}\boldsymbol{N}
\\
&=\bar{\boldsymbol{W}}\boldsymbol{A}_s\boldsymbol{S}_s+\bar{\boldsymbol{W}}\boldsymbol{N}.
\end{aligned}
\label{tuidao}
\end{equation}

In \eqref{tuidao}, $\boldsymbol{Y}_{\mathrm{ZF}}\in \mathbb{C}^{M\times P}$ is the output signal after spatial ZF, which retains only the target signal output and noise. $\boldsymbol{O}_{M \times L}$ denotes a zero matrix with a dimension of $M \times L$. Then, $S$ target directions $(\phi_s, \theta_s), s=L+1, \cdots, L+S $, can be estimated based on $\boldsymbol{Y}_{\mathrm{ZF}}$ using DoA estimation algorithms. Specifically, for MUSIC, the autocorrelation matrix of the received signal is:
\begin{equation}               
\boldsymbol{R}=\frac{\boldsymbol{Y}_{\mathrm{ZF}}\boldsymbol{Y}_{\mathrm{ZF}}^{H}}{P}\in \mathbb{C}^{M\times M}.
    \label{autocorrelation}
\end{equation}

The eigenvalue decomposition of the autocorrelation matrix is obtained as
\begin{equation}
    \boldsymbol{R}=\boldsymbol{Q}_s\boldsymbol{\Lambda}_s \boldsymbol{Q}_{s}^{H}+\boldsymbol{Q}_n\boldsymbol{\Lambda}_n\boldsymbol{Q}_{n}^{H},
\end{equation}
where $\boldsymbol{Q}_{s}\in \mathbb{C}^{M\times S}$, $\boldsymbol{Q}_{n}\in \mathbb{C}^{M\times (M-S)}$ are the eigenvectors spanning the signal and noise subspaces, respectively. $\boldsymbol{\Lambda}_s\in \mathbb{C}^{S\times S}$ and $\boldsymbol{\Lambda}_n\in \mathbb{C}^{(M-S)\times (M-S)}$ are both diagonal matrix composed of the $S$ largest eigenvalues for signal
and the remaining $M-S$ eigenvalues for noise, respectively. Define $\boldsymbol{b}\left( \phi, \theta \right) = \bar{\boldsymbol{W}} \boldsymbol{ \alpha }\left( \phi, \theta \right) \in \mathbb{C}^{M\times 1}$. Based on the algorithm in \cite{oh1993sequential}, the MUSIC pseudo-spectrum can be obtained by leveraging the orthogonality of the signal subspace to the noise subspace:
\begin{equation}
\label{P_MUSIC}
    P_{\mathrm{MUSIC}}\left( \phi, \theta \right) =\frac{1}{\boldsymbol{b}^H\left( \phi, \theta \right) \boldsymbol{Q}_n\boldsymbol{Q}_{n}^{H}\boldsymbol{b}\left( \phi, \theta \right)}.
\end{equation}

By searching the MUSIC pseudo-spectrum, the DoA pairs $\{(\hat{\phi}_s, \hat{\theta}_s)\}_{s=L+1}^{L+S}$ corresponding to $S$ strongest peaks are taken as the targets' DoAs.

\begin{figure}[tbp]
    \centering  
    
    \subfigure[Without spatial zero-forcing]{
    \label{Sparse_MUSIC}    \includegraphics[width=0.49\columnwidth]{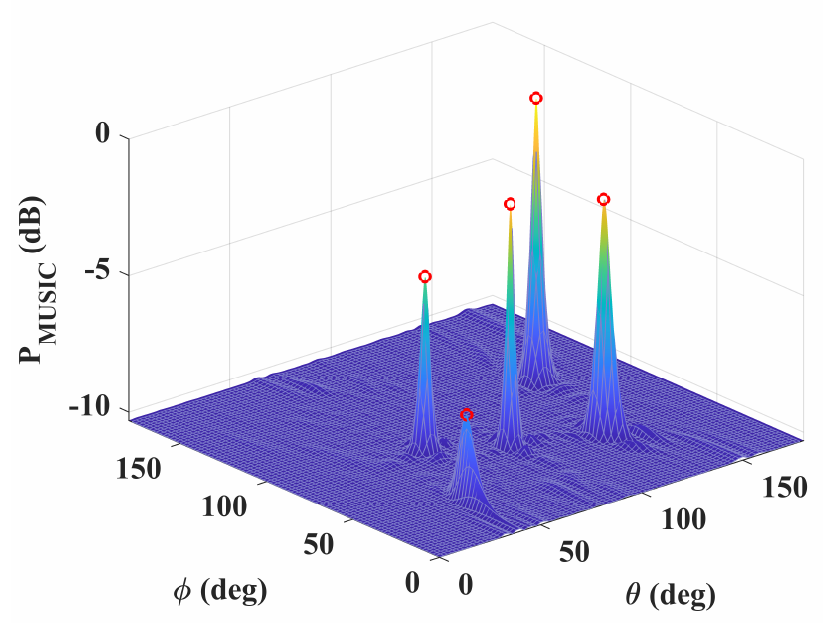}}\subfigure[With CLAM-aided spatial zero-forcing]{
    \label{Sparse_ZF_MUSIC}\includegraphics[width=0.49\columnwidth]{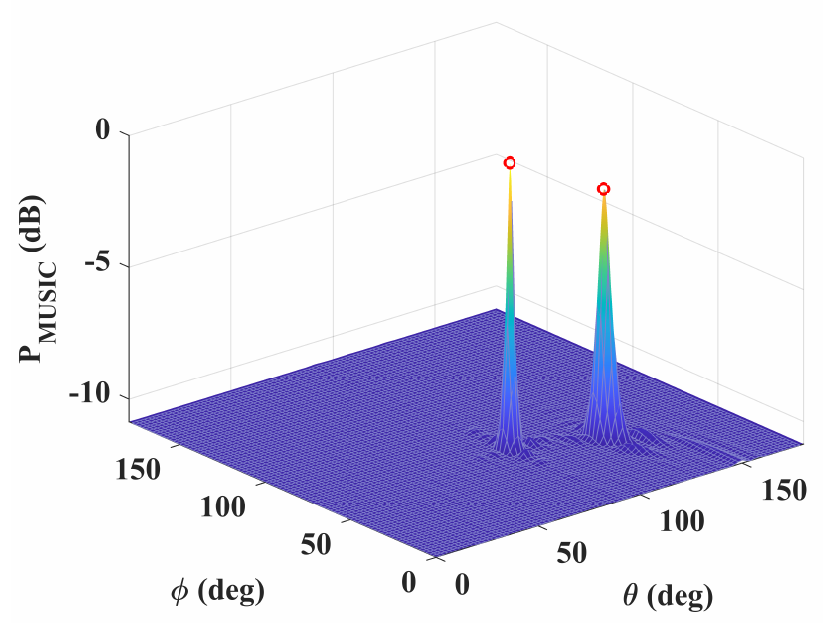}}
    
    \caption{MUSIC spectrum (a) without or (b) with CLAM-aided spatial clutter suppression via zero-forcing.}
    \label{3D_Sparse}
\end{figure}

To observe the impact of spatial ZF on the received signal, we consider a BS equipped with a $32\times 32$ UPA (i.e., $M=1024$ elements). Three clutter signals are from DoAs $(38.7^\circ,\,46^\circ)$, $(84.3^\circ,\,65^\circ)$, and $(111.8^\circ,\,143^\circ)$, while two targets arrive from $(52.4^\circ,\,126^\circ)$ and $(66.8^\circ,\,92^\circ)$. Under this setup, the clutter and target DoAs are well separated in the spatial domain. Fig.\;\ref{Sparse_MUSIC} shows the MUSIC pseudo-spectrum used to estimate the $L$ clutter and $S$ target directions in the OFDM-ISAC system. After applying CLAM-aided spatial suppression, deep nulls are introduced at the clutter DoAs (Fig.\;\ref{Sparse_ZF_MUSIC}), thereby mitigating clutter influence and reducing estimation bias/errors. 

\subsection{CLAM-enabled Joint Spatial-Doppler Domain Clutter Suppression}
\label{C}

\begin{figure}[tbp]
    \centering  
    
    \subfigure[Without spatial zero-forcing]{
    \label{Close_MUSIC}    \includegraphics[width=0.49\columnwidth]{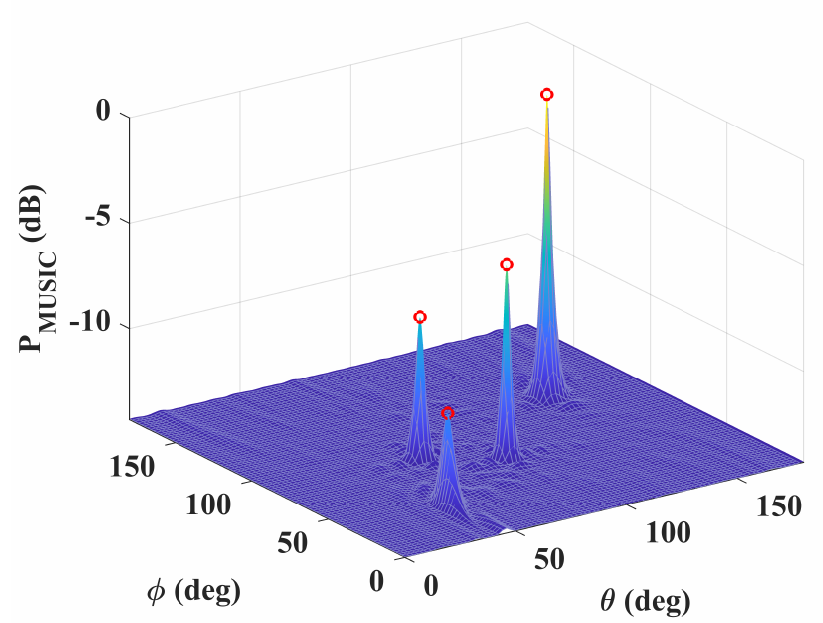}}\subfigure[With CLAM-aided spatial zero-forcing]{
    \label{Close_ZF_MUSIC}\includegraphics[width=0.49\columnwidth]{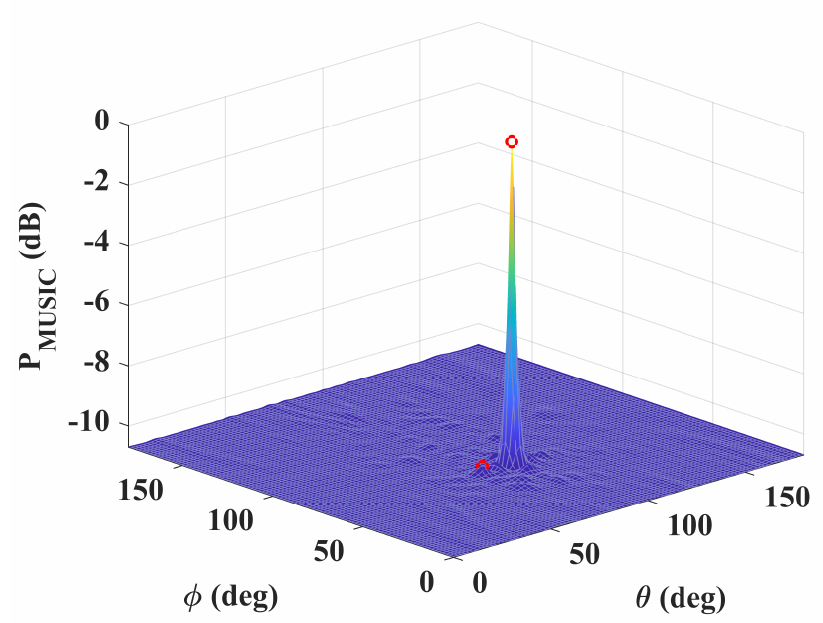}}
    
    \caption{MUSIC spectrum of close DoAs (a) without or (b) with CLAM-aided spatial clutter suppression via zero-forcing.}
    \label{3D_Close}
\end{figure}

When the DoAs of moving targets lie very close to those of clutter scatterers, the aforementioned method encounters challenges. An illustrative example is given in Fig.\;\ref{3D_Close}. Compared with the configuration in Fig.\;\ref{3D_Sparse}, the DoA of one target is changed from $(52.4^\circ,\,126^\circ)$ to $(113.8^\circ,\,141^\circ)$, yielding a minimum separation of $(2^\circ,\,-2^\circ)$ from a clutter signal.

Because a finite array aperture limits angular resolution, Fig.\;\ref{Close_MUSIC} demonstrates that the conventional MUSIC merges the adjacent peaks into a single maximum located between the clutter and target DoAs, so the BS cannot distinguish between the two directions. In Fig.\;\ref{Close_ZF_MUSIC}, applying a CLAM-based ZF projector places a deep null at the clutter direction $(111.8^\circ,\,143^\circ)$ but also suppresses the adjacent target at $(113.8^\circ,\,141^\circ)$. Once the true target is attenuated, MUSIC may lock onto a spurious maximum; in Fig.\;\ref{Close_ZF_MUSIC}, the misestimated target appears near $(65.7^\circ,\,75.9^\circ)$, far from any actual target or clutter direction.

To tackle this challenge, we propose a joint spatial-Doppler domain clutter suppression scheme. By exploiting Doppler difference in addition to spatial ZF, the method discriminates targets from clutter even when their DoAs are close. The core idea for distinguishing targets and clutter in close angular proximity is to concentrate on each estimated target's direction, regarding any other angles as relative \textquoteleft clutter\textquoteright.
\begin{figure}[tbp]
    \centering        \includegraphics[width=0.95\columnwidth]{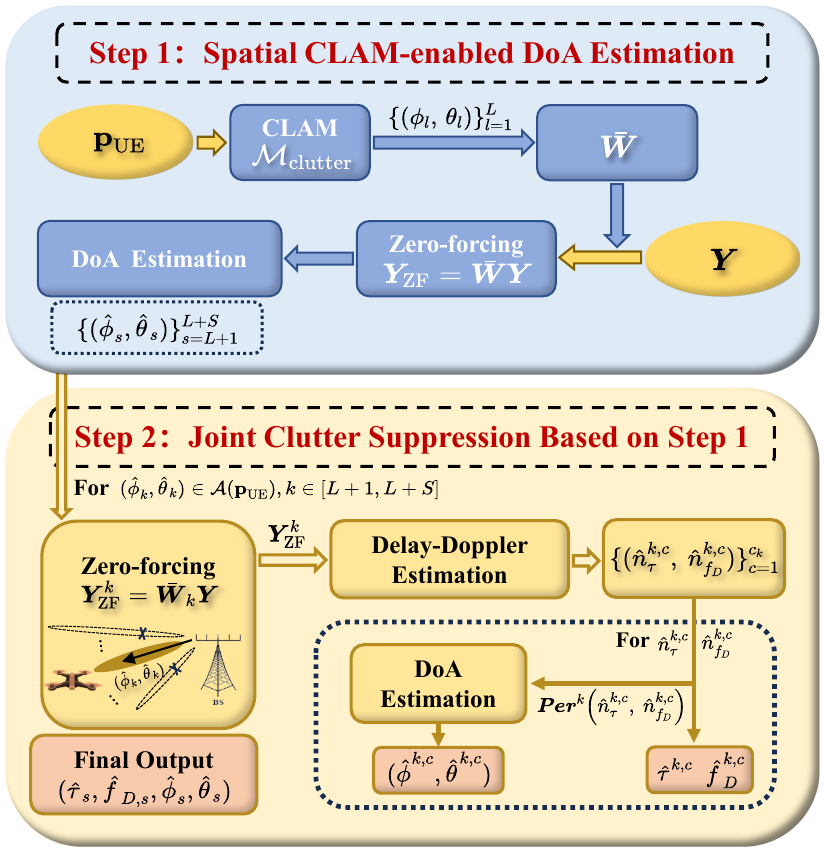}
    \caption{Flowchart of the proposed joint spatial-Doppler clutter suppression algorithm.}
    \label{2-step flowchart}
\end{figure}

The flowchart of the proposed joint spatial-Doppler domain clutter suppression
algorithm is shown in Fig.\;\ref{2-step flowchart}. In the first step, CLAM-aided angle estimation yields the $S$ candidate target DoAs $\{(\hat{\phi}_s,\hat{\theta}_s)\}_{s=L+1}^{L+S}$, some of which may still be deviated relative to the true DoAs. We add these estimated DoAs to the clutter angle set to form a new set:
\begin{equation}               
    \mathcal{A} \left( \mathbf{p}_{\text{UE}} \right) \triangleq \Theta \left( \mathbf{p}_{\text{UE}} \right)\cup\{\hat{\phi}_s, \hat{\theta}_s\}_{s=L+1}^{L+S}.
    \label{new_cluster}
\end{equation}

Here, $\mathcal{A}\big(\mathbf{p}_{\mathrm{UE}}\big)$ collects all clutter DoAs and the estimated target DoAs at the BS for the UE position $\mathbf{p}_{\mathrm{UE}}$. Next, each DoA in $\mathcal{A} \left( \mathbf{p}_{\text{UE}} \right)$ needs to be processed to further distinguish between targets and scatterers from a close direction, or correct any possible errors. So, for the DoA $(\hat{\phi}_k, \hat{\theta}_k) \in \mathcal{A} \left( \mathbf{p}_{\text{UE}} \right)$, which is the specific DoA on which we would like to concentrate in each iteration, all angles in $\mathcal{A} \left( \mathbf{p}_{\text{UE}} \right)$ need to be suppressed except for $(\hat{\phi}_k, \hat{\theta}_k)$ itself. For each round of focusing on each angle, the other angles become relative \textquoteleft clutter\textquoteright, and the array response of the ‘clutter’ $\boldsymbol{C}_k\in \mathbb{C}^{M\times \left(L+S-1\right)}$ can be represented as 
\begin{equation}    \begin{aligned}
    \boldsymbol{C}_k=&\boldsymbol{\alpha} \left( \mathcal{A} \left( \mathbf{p}_{\text{UE}} \right)  \backslash \{(\hat{\phi}_k, \hat{\theta}_k) \} \right)\\
    =&\left[\, \boldsymbol{\alpha }( \phi_1, \theta_1), \cdots, \boldsymbol{\alpha }( \phi_{L}, \theta_{L} ), \boldsymbol{\alpha }( \hat{\phi}_{L+1}, \hat{\theta}_{L+1} ), \cdots, \right. \\
    &\left. \boldsymbol{\alpha }( \hat{\phi}_{k-1} \hat{\theta}_{k-1} ), \boldsymbol{\alpha }( \hat{\phi}_{k+1}, \hat{\theta}_{k+1}), \cdots, \boldsymbol{\alpha }( \hat{\phi}_{L+S}, \hat{\theta}_{L+S} ) \, \right].
\end{aligned}
\end{equation}

Then, $\boldsymbol{W}_k=\boldsymbol{I}-\boldsymbol{C}_k\left( \boldsymbol{C}_k^H\boldsymbol{C}_k \right) ^{-1}\boldsymbol{C}_k^H\in \mathbb{C}^{M\times M}$ can be employed to construct the normalized ZF matrix:
\begin{equation}     
    \bar{\boldsymbol{W}}_k=\boldsymbol{W}_k/||\boldsymbol{W}_k||_F \in \mathbb{C}^{M\times M}.
    \label{beamforming vector}
\end{equation}

With \eqref{beamforming vector}, spatial ZF can be conducted by
\begin{equation}
    \boldsymbol{Y}_{\mathrm{ZF}}^k=\bar{\boldsymbol{W}}_k\boldsymbol{Y}  \in \mathbb{C}^{M \times P}.
    \label{beamforming}
\end{equation}

As described earlier in Fig.\;\ref{3D_Close}, when two directions are closely spaced, their MUSIC peaks may merge into a single maximum. Spatial ZF on this merged peak will suppress the MUSIC spectrum at both angles simultaneously. However, spatial ZF essentially leaves the temporal structure intact; hence, the target’s delay and Doppler are preserved (their values are unaffected by a purely spatial projection). Consequently, the target–clutter ambiguity can then be resolved in the delay–Doppler domain. Moreover, by reducing the number of competing components, the projection step also improves the accuracy of subsequent DoA refinement.

It is worth mentioning that in the presence of adjacent angles, the target directions estimated in Step 1 may not be accurate; consequently, the delay-Doppler processing in Step 2 may face two representative situations, and either of them can benefit from our algorithm:
\begin{itemize}
    \item If the target DoA estimations in Step 1 are correct or only slightly deviated, each iteration suppresses all non-candidate directions, so Step~2 reduces to single-target delay–Doppler estimation. In this setting, efficient matched-filter/FFT-based processing can achieve optimal performance for a single target.
    
    \item If the target DoA estimations in Step 1 have significant errors, as illustrated in Fig.\;\ref{Close_ZF_MUSIC}, the ZF projector will place nulls at directions with no actual signal. The residual signal then contains multiple components; we therefore estimate their delays and Dopplers, use these estimates to discriminate among the components, re-estimate their DoAs using the methods in this subsection, and finally cluster the results to obtain the final estimated sensing parameters, correcting estimation errors in Step 1.
\end{itemize}

\makeatletter
\renewcommand{\alglinenumber}[1]{\scriptsize\the\numexpr#1-1\relax:}
\makeatother

\begin{algorithm}[tbp]
\caption{CLAM-Enabled Joint Spatial-Doppler Domain Clutter Suppression}
\label{algorithm}
\textbf{Input:} $\mathbf{p}_{\text{UE}}$, $\mathcal{M}_{\mathrm{clutter}}$, $\boldsymbol{Y}\in \mathbb{C}^{M\times P}$ \\
\textbf{Output:} Target's delay-Doppler-DoA pair $(\hat{\tau}_s, \hat{f}_{D,s}, \hat{\phi}_s, \hat{\theta}_s)$ \\
\textbf{Step 1: Spatial CLAM-aided Angle Estimation}

\begin{algorithmic}[1] 

\State Obtain $\Theta \left( \mathbf{p}_{\text{UE}} \right)$ from CLAM: $\mathcal{M}_{\mathrm{clutter}}$
\State $ \boldsymbol{C}=\boldsymbol{\alpha }\left( \Theta \left( \mathbf{p}_{\text{UE}} \right) \right), \boldsymbol{W}=\boldsymbol{I}_M-\boldsymbol{C}( \boldsymbol{C}^H\boldsymbol{C} ) ^{-1}\boldsymbol{C}^H$
\State Zero-forcing $\boldsymbol{Y}$ by $\boldsymbol{Y}_{\mathrm{ZF}}=\bar{\boldsymbol{W}}\boldsymbol{Y}$
\State Calculate the MUSIC spectrum $P_{\text{MUSIC}}\left( \phi, \theta \right)$ 
\State Search the spectrum to obtain $\{(\hat{\phi}_s, \hat{\theta}_s)\}_{s=L+1}^{L+S}$ corresponding to the $S$ peaks 
\end{algorithmic}

\textbf{Step 2: Joint Clutter Suppression Based on Step 1}
\begin{algorithmic}[1] 
\State $\mathcal{A} \left( \mathbf{p}_{\text{UE}} \right) \triangleq \Theta \left( \mathbf{p}_{\text{UE}} \right)\cup\{\hat{\phi}_s, \hat{\theta}_s\}_{s=L+1}^{L+S}$
\For {each $(\hat{\phi}_k, \hat{\theta}_k) \in \mathcal{A} \left( \mathbf{p}_{\text{UE}} \right)$ }
    \State $\boldsymbol{C}_k=\boldsymbol{\alpha} \left( \mathcal{A} \left( \mathbf{p}_{\text{UE}} \right)  \backslash \{(\hat{\phi}_k, \hat{\theta}_k) \} \right)\in \mathbb{C}^{M\times \left(L+S-1\right)} $
    \State $\boldsymbol{W}_k=\boldsymbol{I}-\boldsymbol{C}_k\left( \boldsymbol{C}_k^H\boldsymbol{C}_k \right) ^{-1}\boldsymbol{C}_k^H\in \mathbb{C}^{M\times M}$
    \State Normalized ZF matrix $\bar{\boldsymbol{W}}_k=\boldsymbol{W}_k/||\boldsymbol{W}_k||_F$
    
    \State Perform spatial ZF: $\boldsymbol{Y}_{\mathrm{ZF}}^k= \bar{\boldsymbol{W}}_k\boldsymbol{Y}$
    \State Tensor reshape by $\boldsymbol{F}_{\mathrm{ZF}}^{k}(:, n, \gamma) = \boldsymbol{Y}_{\mathrm{ZF}}^k(:, p_{n,\gamma})$
    \State Obtain $\boldsymbol{Per}^k$ by delay-Doppler processing on $\boldsymbol{F}_{\mathrm{ZF}}^{k}$
    \State Find the maximum pairs $\{ (\hat{n}_{\tau}^{k,c},\ \hat{n}_{f_D}^{k,c})\}_{c=1}^{c_k}$ in $\boldsymbol{Per}^k$

    \For{$(\hat{n}_{\tau}^{k,c},\ \hat{n}_{f_D}^{k,c})$}
        \State Delay $\hat{\tau}^{k,c}$ and Doppler $\hat{f}_D^{k,c}$ calculation
        \State Signal extraction with $\boldsymbol{Per}^k\left( \hat{n}_{\tau}^{k,c},\ \hat{n}_{f_D}^{k,c} \right)$
        \State FFT/MUSIC DoA estimation
    \EndFor    
    
    \State Obtain $c_k$ DoA pairs: $\{(\hat{\phi}^{k, c}, \hat{\theta}^{k, c})\}_{c=1}^{c_k}$
        
\EndFor
\State Cluster to match the final sensing parameters of targets.
\end{algorithmic}
\end{algorithm}

To perform 2-D FFT for delay–Doppler estimation, we need to reshape $\boldsymbol{Y}_{\mathrm{ZF}}^k$ back into a 3-D tensor by
\begin{equation}
    \boldsymbol{F}_{\mathrm{ZF}}^{k}(:, n, \gamma) = \boldsymbol{Y}_{\mathrm{ZF}}^k(:, p_{n,\gamma}),
\end{equation}
where $ \boldsymbol{F}_{\mathrm{ZF}}^{k} \in \mathbb{C}^{M \times N_{\mathrm{sc}} \times N_{\mathrm{sym}}} $, $n=\mathrm{mod}_{N_{sc}}(p_{n,\gamma})$ and $\gamma=\left\lfloor p_{n,\gamma}/N_{sc} \right\rfloor$, $0 \leq n \leq N_{\mathrm{sc}}-1,\,0 \leq \gamma \leq N_{\mathrm{sym}}-1$.
Next, $N_{\tau}^{\rm{IFFT}}$-point IFFTs and $N_{f_D}^{\rm{FFT}}$-point FFTs are performed on $\boldsymbol{F}_{\mathrm{{ZF}}}^{k}$ in the subcarrier and symbol domains. Typically, to enhance resolution, oversampling techniques are employed during IFFT and FFT, that is, $N_{\tau}^{\rm{IFFT}} > N_{\mathrm{sc}}$ and $N_{f_D}^{\rm{IFFT}} > N_{\mathrm{sym}}$. Therefore, we use zero-padding to convert $\boldsymbol{F}_{\mathrm{ZF}}^{k} \in \mathbb{C}^{M \times N_{\mathrm{sc}} \times N_{\mathrm{sym}}}$ into $\boldsymbol{F}_{\mathrm{ZF,ZP}}^{k} \in \mathbb{C}^{M \times N_{\tau}^{\rm{IFFT}} \times N_{f_D}^{\rm{FFT}}}$. Then, the resulting delay-Doppler spectrum $\boldsymbol{Per}^k \in \mathbb{C}^{M \times N_{\tau}^{\rm{IFFT}} \times N_{f_D}^{\rm{FFT}}}$ after spatial ZF can be obtained by employing the 2D-periodogram algorithm in \cite{dai2025tutorial}:
\begin{equation}
\label{2D-FFT}
\begin{aligned}
    \boldsymbol{Per}^k &= \Big|\text{FFT}^3_{N_{f_D}^{\rm{FFT}}}\left( \text{IFFT}^2_{N_{\tau}^{\rm{IFFT}}}\left( \boldsymbol{F}_{\mathrm{ZF,ZP}}^{k} \right) \right) \Big|^2 \\
    &=\frac{1}{N_{\mathrm{sc}}N_{\mathrm{sym}}} \Big| \boldsymbol{\mathrm{W}}_{N_{\tau}^{\rm{IFFT}}} \boldsymbol{F}_{\mathrm{ZF,ZP}}^{k} \boldsymbol{\mathrm{W}}^*_{N_{f_D}^{\rm{FFT}}}  \Big|^2,
\end{aligned}    
\end{equation}
where $\mathrm{FFT}_N^n(\cdot)$ and $\mathrm{IFFT}^n_N(\cdot)$ represent $N$-point FFT and IFFT on the $n^{th}$ dimension. $\boldsymbol{\mathrm{W}}_N$ is the $N$-point IDFT matrix, which is detailed in \cite{dai2025tutorial}.

Next, we traverse the delay–Doppler spectrum using a maximum search. When a sharp peak is accompanied by elevated neighboring samples, we group the peak and its surrounding high-magnitude region into a single class. Assuming that $c_k$ classes are identified in the $k^{th}$ iteration, then the corresponding peaks $\{ (\hat{n}_{\tau}^{k,c},\ \hat{n}_{f_D}^{k,c})\}_{c=1}^{c_k}$ are taken as the estimated delay-Doppler indices. Then, for each $(\hat{n}_{\tau}^{k,c},\ \hat{n}_{f_D}^{k,c})$, the estimated delay and Doppler can be calculated by:
\begin{equation}
\label{delay_doppler}
    \hat{\tau}^{k,c}=\frac{\hat{n}_{\tau}^{k,c}}{\Delta f N_{\tau}^{\mathrm{IFFT}}},\,\,\, \hat{f}_D^{k,c}=\frac{\hat{n}_{f_D}^{k,c}}{T_s N_{f_D}^{\mathrm{FFT}}}.
\end{equation}

By extracting the spatial domain signals at each of the estimated delay–Doppler indices, clutter suppression in the joint spatial-Doppler domain can be achieved. Angle domain ZF and delay domain signal extraction belong to the spatial domain part, while Doppler domain extraction can be referred to as clutter suppression in the Doppler domain. Then, based on $\boldsymbol{Per}^k( \hat{n}_{\tau}^{k,c},\ \hat{n}_{f_D}^{k,c} )$, DoA refinement can be performed with the periodogram/FFT method or super-resolution algorithms such as MUSIC, treating other residual components as interference. As a result, $\{(\hat{\phi}^{k, c}, \hat{\theta}^{k, c})\}_{c=1}^{c_k}$ can be obtained. For the $k^{th}$ iteration, $c_k$ targets are detected, and thus Step 2 will estimate $\sum_{k=1}^{S} c_k$ DoAs in total. Using the clustering matching algorithm, the final DoAs of $S$ targets can be obtained.

The proposed CLAM-enabled joint spatial-Doppler domain clutter suppression method is summarized in Algorithm \ref{algorithm}.

\subsection{Computational Complexity}
\label{Complexity}
We measure computational complexity by counting the number of complex multiplications. Specifically, multiplying $\boldsymbol{A}_{M\times N}\boldsymbol{B}_{N\times P}=\boldsymbol{C}_{M\times P}$ costs $MNP$ multiplications; the eigendecomposition of an $N \times N$ matrix costs $N^3$; and we likewise count inversion of an $N \times N$ matrix as $N^3$. Hence, constructing the ZF projector
\[
\boldsymbol{W}=\boldsymbol{I}_M-\boldsymbol{C}\big(\boldsymbol{C}^H\boldsymbol{C}\big)^{-1}\boldsymbol{C}^H\in\mathbb{C}^{M\times M}
\] 
requires $K_{\phi,\theta}\,(M+K_{\phi,\theta})^2$ complex multiplications, where $K_{\phi,\theta}$ is the number of directions being zero-forced. Let $r_{\phi}$ and $r_{\theta}$ denote the azimuth/zenith grid sizes for MUSIC scanning; e.g., dividing $[0^\circ,180^\circ]$ with $0.1^\circ$ steps gives $r_{\phi}=r_{\theta}=1801$.

\begin{table*}[t!]
    \centering
    \caption{\textbf{Computational complexity (in complex multiplications)}}
    \label{tab:complexity}
    \setcellgapes{1.5pt}
    \makegapedcells
    \begin{tabular}{|c|c|}
    \hline
    \textbf{Algorithm} & \textbf{Complexity} \\
    \hline
    FFT & $N_{\rm sym}N_{\rm sc}N^{\rm FFT}\!\times\!\Big(1+\tfrac{1}{2}\log_2 N^{\rm FFT}\Big)$ \\
    \hline
    Spatial CLAM-aided FFT & $N_{\rm sym}N_{\rm sc}N^{\rm FFT}\!\times\!\Big(1+\tfrac{1}{2}\log_2 N^{\rm FFT}\Big)+N_{\rm sym}N_{\rm sc}M^2$ \\
    \hline
    Joint CLAM-aided FFT & \makecell{$S\!\times\!\Big[\,(L\!+\!S\!-\!1)(M\!+\!L\!+\!S\!-\!1)^2+N_{\rm sym}N_{\rm sc}M^2+MN_{\rm sym}\tfrac{1}{2}N_{\tau}^{\rm IFFT}\log_2 N_{\tau}^{\rm IFFT}$ \\ $+\,MN_{\tau}^{\rm IFFT}\tfrac{1}{2}N_{f_D}^{\rm FFT}\log_2 N_{f_D}^{\rm FFT}+MN_{\tau}^{\rm IFFT}N_{f_D}^{\rm FFT}+\tfrac{1}{2}N^{\rm FFT}\log_2 N^{\rm FFT}+N^{\rm FFT}\,\Big]$} \\
    \hline
    MUSIC & $N_{\rm sym}N_{\rm sc}M^2+M^3+r_{\phi}r_{\theta}(M+1)\big(M-(L+S)\big)$ \\
    \hline
    Spatial CLAM-aided MUSIC & $2N_{\rm sym}N_{\rm sc}M^2+M^3+r_{\phi}r_{\theta}(M+1)\big(M-S\big)$ \\
    \hline
    Sequential ZF–MUSIC & $M^2(L\!+\!S)\big(2N_{\rm sym}N_{\rm sc}+M+r_{\phi}r_{\theta}\big)+\sum_{k_{\phi,\theta}=1}^{L+S}\! 2k_{\phi,\theta}\big(M+k_{\phi,\theta}\big)^2$ \\
    \hline
    Joint CLAM-aided MUSIC & \makecell{$S\!\times\!\Big[\,(L\!+\!S\!-\!1)(L\!+\!S\!-\!1\!+\!M)^2+N_{\rm sym}N_{\rm sc}M^2+MN_{\rm sym}\tfrac{1}{2}N_{\tau}^{\rm IFFT}\log_2 N_{\tau}^{\rm IFFT}$ \\ $+\,MN_{\tau}^{\rm IFFT}\tfrac{1}{2}N_{f_D}^{\rm FFT}\log_2 N_{f_D}^{\rm FFT}+MN_{\tau}^{\rm IFFT}N_{f_D}^{\rm FFT}+M^2+M^3+r_{\phi}r_{\theta}\big(M^2-1\big)\,\Big]$} \\
    \hline
    \end{tabular}
\end{table*}

For FFT-based methods, performing an $N^{\rm FFT}$-point FFT on the demodulated, symbol-aligned data matrix $\boldsymbol{F}_{M\times (N_{\rm sym}N_{\rm sc})}$ costs $N_{\rm sym}N_{\rm sc}\times \tfrac{1}{2}N^{\rm FFT}\log_2 N^{\rm FFT}$. Computing per-bin energy adds $N_{\rm sym}N_{\rm sc}N^{\rm FFT}$ multiplications. Applying spatial ZF at clutter angles is equivalent to left-multiplying the $M\times P$ snapshot matrix by $\boldsymbol{W}_{M\times M}$, incurring $N_{\rm sym}N_{\rm sc}M^2$ multiplications. For joint spatial–Doppler suppression, forming $\boldsymbol{W}$ and projecting the data for each of the $S$ candidate DoAs cost $(L\!+\!S\!-\!1)(M\!+\!L\!+\!S\!-\!1)^2$ and $N_{\rm sym}N_{\rm sc}M^2$, respectively. Delay–Doppler processing then requires $MN_{\rm sym}\tfrac{1}{2}N_{\tau}^{\rm IFFT}\log_2 N_{\tau}^{\rm IFFT}$ for the delay IFFT and $MN_{\tau}^{\rm IFFT}\tfrac{1}{2}N_{f_D}^{\rm FFT}\log_2 N_{f_D}^{\rm FFT}$ for the Doppler FFT; computing the 2D periodogram adds $MN_{\tau}^{\rm IFFT}N_{f_D}^{\rm FFT}$ multiplications. A conventional FFT-based DoA estimator can then be applied.

For MUSIC-based methods, forming the covariance of an $M\times (N_{\rm sym}N_{\rm sc})$ matrix costs $N_{\rm sym}N_{\rm sc}M^2$. For each grid point $(\phi,\theta)$ in \eqref{P_MUSIC}, evaluating the denominator via $\boldsymbol{b}^H(\phi,\theta)\boldsymbol{Q}_n$ followed by an inner product costs $(M+1)\big(M-K_{\phi,\theta}\big)$ multiplications—considerably fewer than full sequential matrix products. In path-by-path sequential ZF–MUSIC\cite{oh1993sequential, Zhou2024Single}, which we introduce as a stronger baseline algorithm, each round estimates $k_{\phi,\theta}\in\{1,2,\ldots, L\!+\!S\}$ directions; after each round, we (re)build the ZF projector and re-project the data, giving the cumulative cost $\sum_{k_{\phi,\theta}=1}^{L+S}\!\big[2k_{\phi,\theta}(M+k_{\phi,\theta})^2+N_{\rm sym}N_{\rm sc}M^2\big]$ across varying projector dimensions. Joint spatial–Doppler suppression shares the same front end for FFT and MUSIC; only the final angle estimator differs.

\begin{figure}[tbp]
    \centering
    \includegraphics[width=0.9\columnwidth]{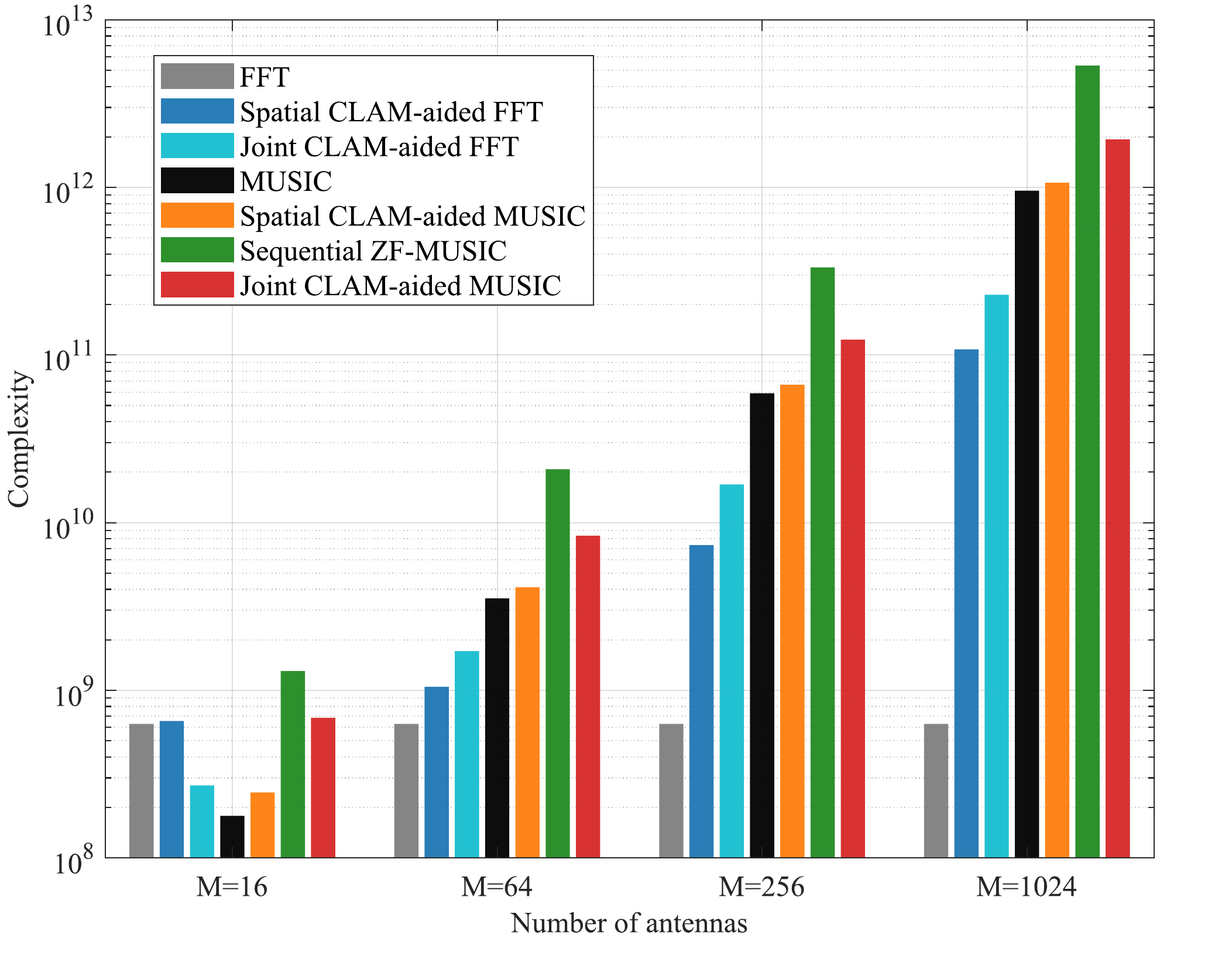}
    \caption{Computational complexity versus array size.}
    \label{fig:complexity}
\end{figure}

Complete expressions are listed in Table~\ref{tab:complexity}. We set $N_{\rm sym}=100$, $N_{\rm sc}=N^{\rm FFT}=N_{\tau}^{\rm IFFT}=N_{f_D}^{\rm FFT}=1024$, $S=2$, $L=3$, $r_{\phi}=r_{\theta}=901$, and $M\in\{4\times4,\,8\times8,\,16\times16,\,32\times32\}$. Fig.\;\ref{fig:complexity} reports the numerical complexity for each algorithm. With small arrays, the large numbers of symbols and subcarriers in OFDM force FFT-based approaches to process many snapshots, so their cost can exceed that of super-resolution methods. As $M$ grows, the $r_{\phi}r_{\theta}M^2$ term drives a steep increase for MUSIC-based algorithms, whereas FFT-based counterparts grow more moderately.

Within each family, spatial CLAM-aided variants add only $N_{\rm sym}N_{\rm sc}M^2$ relative to their conventional versions, making them computationally economical. Sequential ZF–MUSIC is the most expensive because it performs $L\!+\!S$ estimation rounds, each followed by projector updates and re-projection. The joint CLAM-aided algorithms are also costly—each of the $S$ rounds includes a full delay–Doppler stage—but after joint domain suppression and per-bin extraction, the effective data dimension can shrink dramatically (often down to an $M$-length spatial snapshot), limiting subsequent costs. Thus, aided by CLAM priors, the joint scheme attains high clutter suppression performance while remaining practical and cost-effective.

\section{numerical simulation}
\label{numerical simulation}
This section presents simulation results for the proposed algorithms alongside relevant baselines. Over multiple Monte Carlo trials, we evaluate (i) DoA estimation performance with and without Doppler-based clutter suppression, (ii) DoA estimation performance under spatial-only and joint spatial-Doppler clutter suppression, and (iii) target localization accuracy. The key simulation parameters are summarized in Table~\ref{parameters}.
\begin{table}[ht]
    \caption{\textbf{Simulation parameters}}
    \centering 
    \begin{tabular}{cc} 
    \toprule
    Parameter & Setting and Description \\
    \midrule
    Carrier frequency $f_c$ & 28 GHz \\
    Subcarrier spacing $\Delta f$ & 30 kHz \\
    Number of subcarriers $N_{\mathrm{sc}}$ & 1024 \\
    Number of symbols $N_{\mathrm{sym}}$ & 100 \\
    Bandwidth $B$ & 30.72 MHz \\
    FFT size $N^{\mathrm{FFT}}$ & 1024$\times$5 \\
    CP length & 288 \\
    Antenna spacing $d$ & $\lambda/2$ \\
    Number of antennas $M = M_x \times M_z$ & $32 \times 32$ \\
    
    \bottomrule
    \end{tabular}
    \label{parameters}
\end{table}

Assume the range of both azimuth and zenith angle is $[0^\circ,\,180^\circ]$. The BS is located at $(0,\,1)\,\mathrm{km}$ in the horizontal plane. The UE and two clutter scatterers are placed at $(0.1,\,0)\,\mathrm{km}$, $(1,\,0.2)\,\mathrm{km}$, and $(-0.4,\,0)\,\mathrm{km}$, respectively. Target~1 and Target~2 are at $(0.5,\,0.35)\,\mathrm{km}$ and $(0.3,\,0.3)\,\mathrm{km}$, representing a fast- and a slow-moving target, respectively. Target~3 is constructed by applying a small angular perturbation to one clutter direction to create a near-angle case.


The Doppler shift of target is
$f_{D,s}=\tilde{v}_s/\lambda$, where $\tilde{v}$ represents the bistatic radial velocity in \cite{xiao2024integrated}. The specific time delays, Doppler shifts, azimuth angles, and zenith angles for each clutter and low-altitude target are summarized in Table~\ref{parameters}. This configuration jointly evaluates quasi-static clutter, dynamic clutter, slow/fast targets, and close-angle scenarios.

\begin{table}[ht]
    \centering
    \caption{ \textbf{To-be-sensed parameters}}
    \label{Parameters}
    \setcellgapes{3pt} 
    \makegapedcells 
    
    \begin{tabular}{|c|c|c|c|c|}
    \hline
    \textbf{Scatterers} &
    {$\tau \ (\mathrm{\mu s})$} &
    {${f_D} \ (\mathrm{Hz})$} &
    {${\phi\ (\circ)}$} &
    {${\theta\ (\circ)}$}
    \\
    \hline
        Clutter 1 & 3.35 & 186.7 & 84.3 & 65 
        \\
        \hline
        Clutter 2 & 7.34 & 466.7 & 38.7 & 46
        \\
        \hline
        Clutter 3  & 5.25 & 746.7 & \textbf{111.8} & \textbf{143} 
        \\
        \hline
        Target 1 & 4.50 & \textbf{3733.3} & 52.4 & 126
        \\
        \hline Target 2
         & 3.74 & \textbf{373.3} & 66.8 & 92
        \\
        \hline
        Target 3 
        & 3.74 & 373.3 & \textbf{113.8} & \textbf{141}
        \\
        \hline
    \end{tabular}
\end{table}

\subsection{DoAs Estimation with Doppler Domain Clutter Suppression}
\label{MTI-sim}
\begin{figure}[tbp]
    \centering
    \includegraphics[width=0.85\columnwidth]{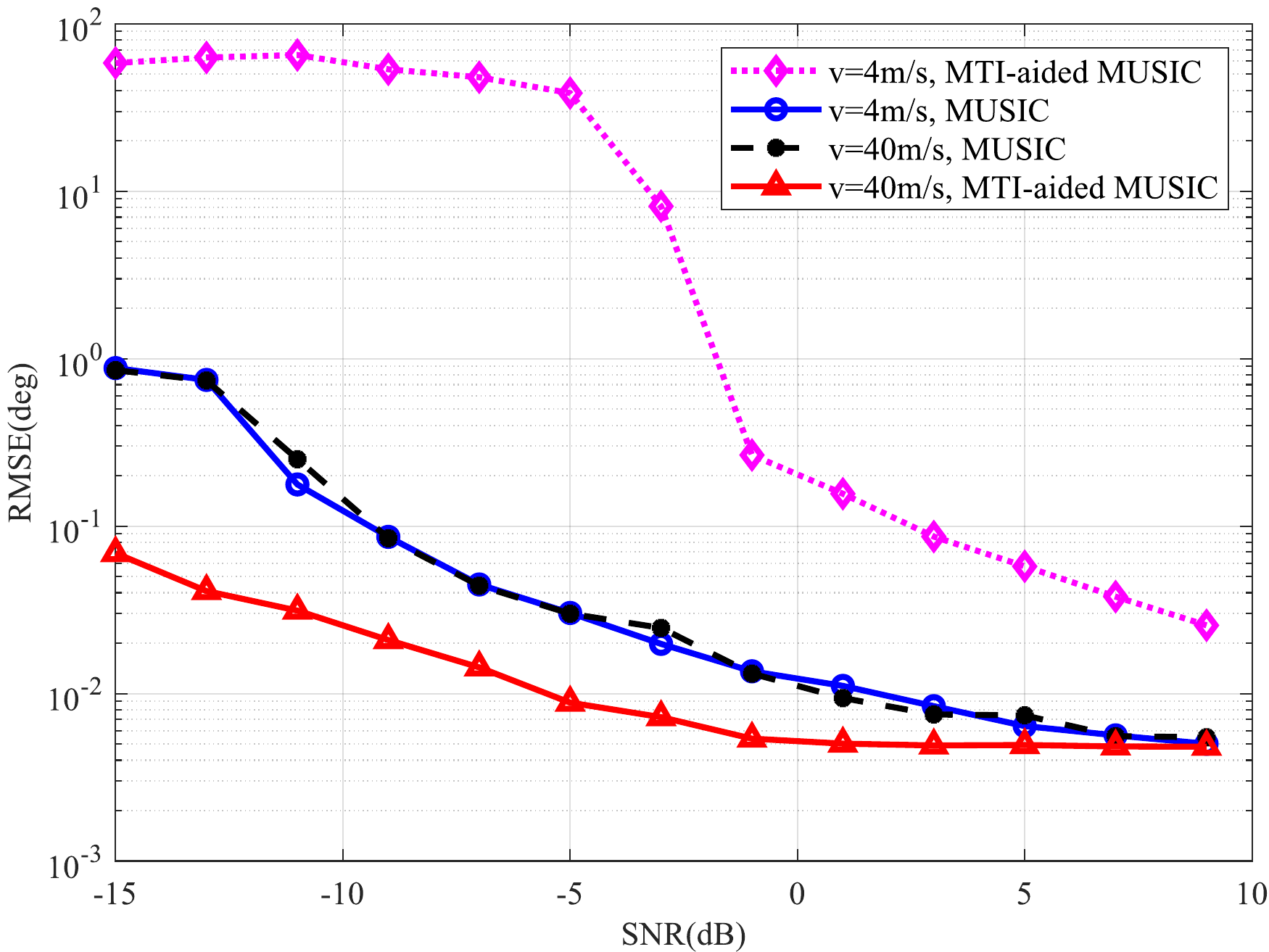}
    \caption{Azimuth-angle RMSE with and without Doppler-based MTI clutter suppression for fast (40\,m/s) and slow (4\,m/s) targets.}
    \label{fig:vs_mti}
\end{figure}

We evaluate the impact of the proposed OFDM MTI method (Sec.~\ref{MTI+OFDM}) to quantify how target speed affects sensing performance. Two independent cases are considered: a fast-moving target (Target~1) with bistatic radial velocity $\tilde{v}=40\,\mathrm{m/s}$ relative to the BS in case~1, and a slow-moving target (Target~2) with $\tilde{v}=4\,\mathrm{m/s}$ in case~2.

Fig.\;\ref{fig:vs_mti} reports the azimuth RMSE obtained (i) by applying MUSIC directly to the uplink measurements and (ii) after first-order MTI clutter suppression as in \eqref{OFDM_MTI}. For OFDM with Doppler-based MTI, high-speed targets achieve noticeably better DoA accuracy. In contrast, for slow-moving targets, even super-resolution processing (MTI-aided MUSIC) may yield poor accuracy and, at low SNR, large estimation errors. This occurs because small Doppler shifts lie within or near the MTI notch, attenuating the desired signal and reducing post-MTI SNR, as analyzed in Sec.~\ref{Limitations}.

\subsection{Widely Spaced DoA Estimation with Spatial Clutter Suppression}

We validate the benefit of CLAM-enabled spatial clutter suppression for low-altitude DoA estimation in a sparse-DoA scene, where Clutter~1–3 and Targets~1–2 are well separated in angle.

\begin{figure}[tbp]
    \centering
    \includegraphics[width=0.8\columnwidth]{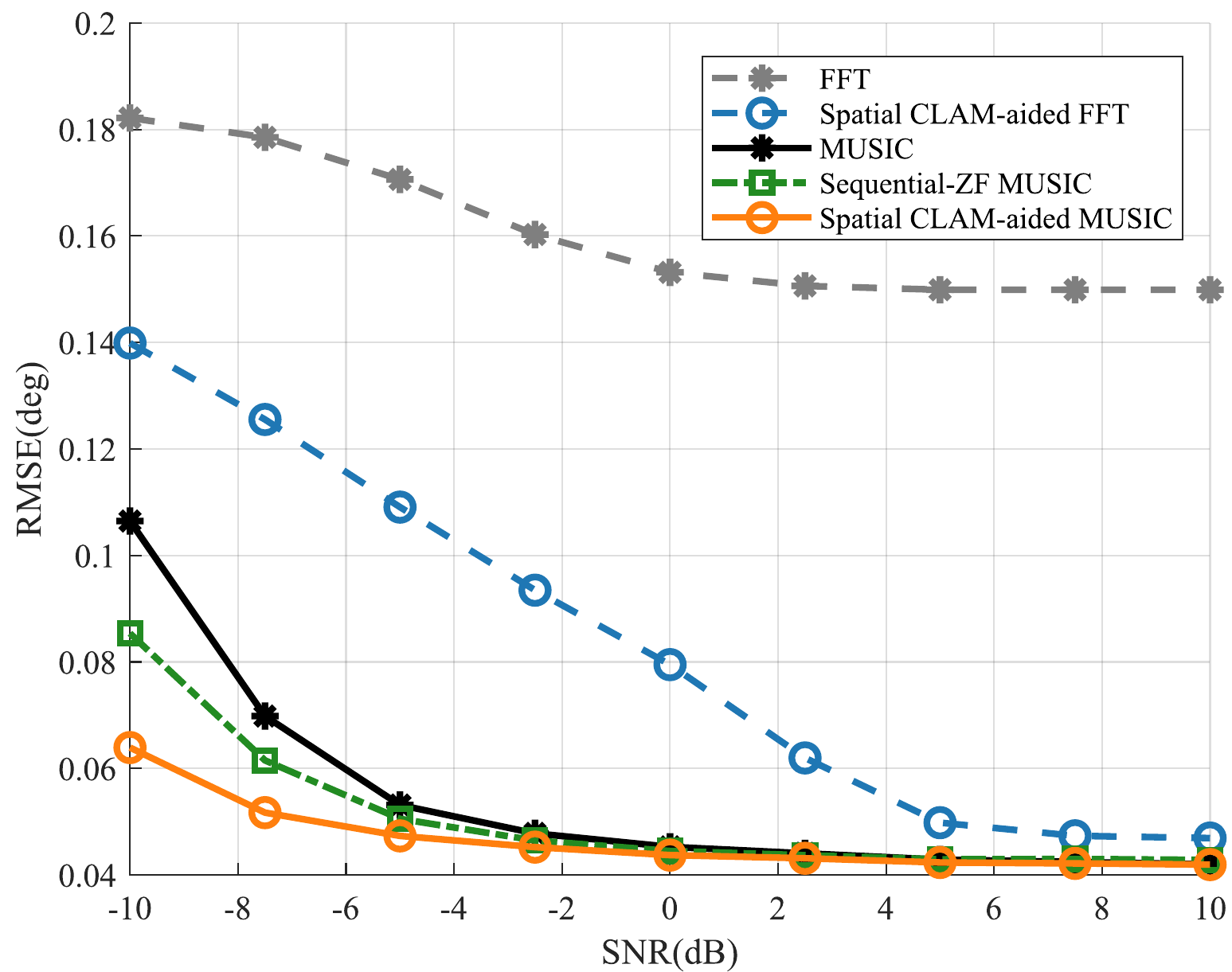}
    \caption{Azimuth-angle RMSE comparison in a sparse-DoA scene.}
    \label{fft}
\end{figure}

Fig.\;\ref{fft} evaluates the method in Sec.~\ref{CLAM-based Spatial Clutter Suppression}. We compare azimuth RMSE versus SNR for: (i) conventional FFT, (ii) spatial-smoothing MUSIC, (iii) path-by-path sequential ZF MUSIC, and (iv) CLAM-aided ZF followed by MUSIC or FFT. Standard MUSIC lacks explicit clutter-angle cancellation. The sequential ZF–MUSIC baseline mentioned in Sec.~\ref{Complexity} iteratively (a) estimates the strongest DoA from the MUSIC spectrum, (b) places a null at all directions estimated, and (c) repeats; while it can reduce the number of active paths per round, early misestimates propagate to later rounds and the iterative procedure is computationally heavy. Under the present settings, CLAM-aided MUSIC/FFT achieves lower RMSE than conventional MUSIC/FFT, and CLAM-aided MUSIC also outperforms the sequential ZF–MUSIC baseline.

As SNR increases, the performance of the three MUSIC-based curves improve and their gap narrows beyond the SNR of 0\,dB, reflecting their shared super-resolution limit for the given array and scenario. In contrast, conventional FFT exhibits a clutter-induced peak-selection bias that does not vanish at high SNR, so its RMSE does not converge to the same minimum as CLAM-aided FFT. By exploiting clutter-angle priors, CLAM removes nearby-clutter bias and enables FFT to approach the accuracy of super-resolution methods.

\subsection{Closely Spaced DoA Estimation with Joint Spatial–Doppler Domain Clutter Suppression}

\begin{figure}[tbp]
    \centering
    \includegraphics[width=0.85\columnwidth]{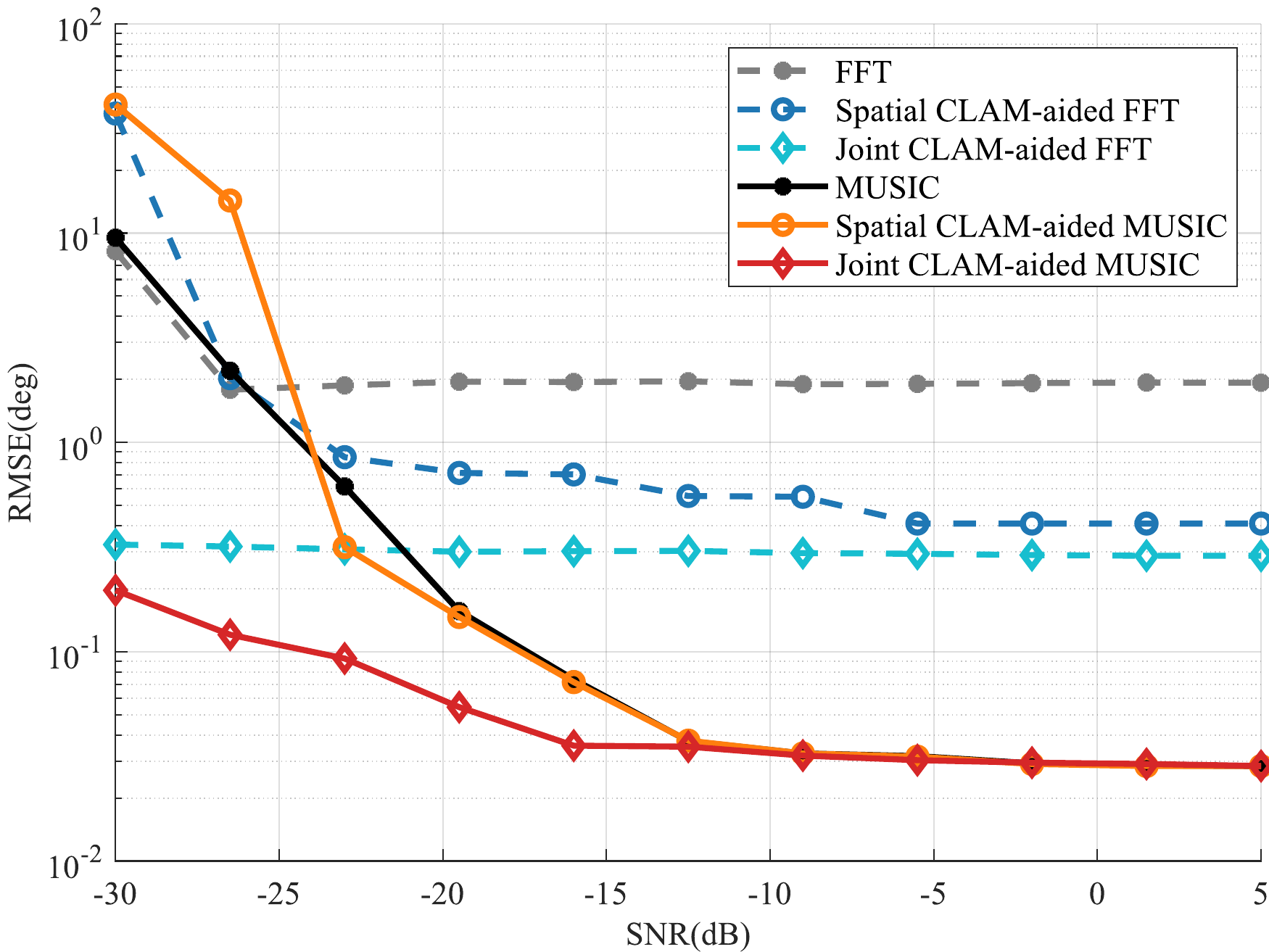}
    \caption{Azimuth-angle RMSE (dB scale) for closely spaced DoAs.}
    \label{2step_M32}
\end{figure}

We now replace Target~2 with Target~3, which lies close to Clutter~3 in angle. Relative to Fig.\;\ref{fft}, Fig.\;\ref{2step_M32} reduces the minimum azimuth/zenith separation between a target and clutter to $(2^\circ,\,-2^\circ)$. The CLAM-enabled \emph{joint} spatial–Doppler scheme in Sec.~\ref{C} avoids relying on angular ZF alone and mitigates interference between adjacent directions, thereby improving DoA accuracy. First, the Step~1 spatial ZF reduces the number of active components, facilitating more accurate delay–Doppler estimation. Second, the improved delay and Doppler estimation for the surviving components enable cleaner target extraction in Step~2.

In multi-target/multi-clutter scenarios, FFT-based methods yield poorer angular accuracy than super-resolution methods (e.g., MUSIC). However, at low SNR, the signal–noise subspace orthogonality degrades, producing erratic peaks; consequently, both conventional MUSIC and spatial CLAM-aided MUSIC can perform poorly.

Within each method family, conventional FFT/MUSIC performs worst. Spatial CLAM-aided FFT/MUSIC is intermediate: its performance degrades at low SNR when purely spatial ZF suppresses energy for adjacent angles. The proposed joint CLAM-aided FFT/MUSIC, which suppresses clutter in the spatial–Doppler domain and processes as few components as possible per iteration with delay–Doppler gating and DoA re-estimation, delivers the best DoA accuracy among the tested approaches.

\subsection{Polar-Coordinate Estimation with or without CLAM}
\begin{figure}[tbp]
    \centering
    \includegraphics[width=0.9\columnwidth]{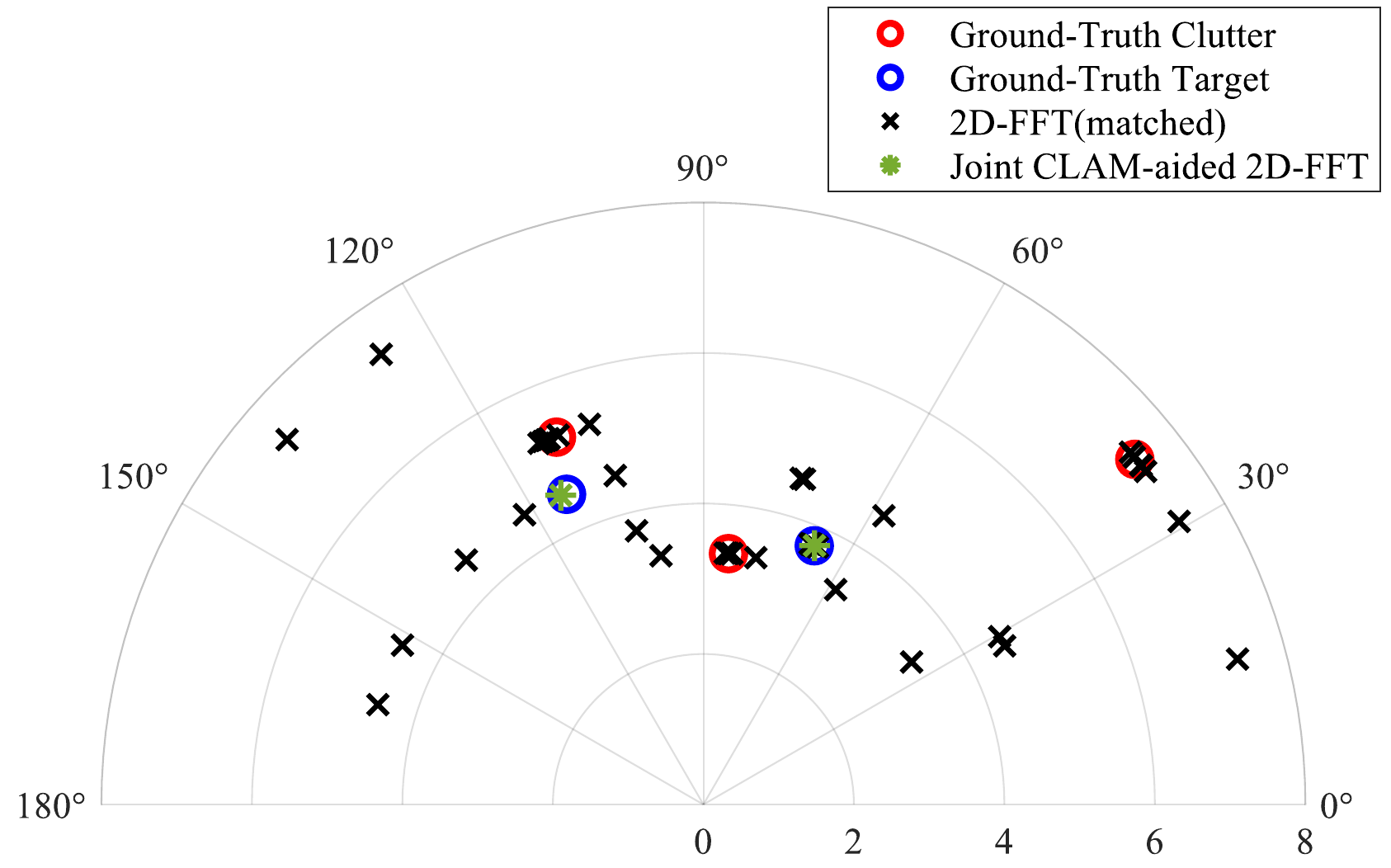}
    \caption{Delay–azimuth estimates with and without joint CLAM-aided clutter suppression over 10 independent runs.}
    \label{semicircle}
\end{figure}

With each target’s estimated delay and DoA, we plot its delay–azimuth location in polar coordinates in Fig.\;\ref{semicircle} as a metric for position-estimation accuracy: the angle denotes azimuth (degrees) and the radius denotes delay (in $\mu$s). The BS is placed at the origin. Note that the plotted delay corresponds to the \emph{bistatic} UE–target–BS path delay, rather than the direct BS–target distance. Given a target’s delay and azimuth, and using the property that points with constant sum of distances to the UE and BS lie on an ellipse (with foci at the UE and BS), the target’s 2D position is determined by the intersection of that ellipse and the BS azimuth ray.

Fig.\;\ref{semicircle} shows results with and without joint CLAM-aided clutter suppression over 10 independent simulations at $-30$\,dB SNR. Because delay–Doppler and angle are estimated separately, each 2D-FFT delay estimate is associated with the nearest ground-truth target and the residual error is retained. Conventional methods frequently produce erroneous estimates at low SNR and exhibit run-to-run bias for both targets and clutter. In contrast, the proposed joint spatial–Doppler suppression yields consistent, accurate estimates across all runs and separates targets from clutter directly. Finally, the poorer localization performance of plain 2D-FFT arises primarily from angular deviations, whereas delay estimates remain comparatively accurate even at low SNR.

\section{Conclusion}
\label{Conclusion}
In this work, we develop CLAM-enabled clutter suppression algorithms for low-altitude UAV ISAC. With clutter angles estimated offline and stored in CLAM, the online processor imposes spatial nulls toward those directions, thereby attenuating clutter while preserving target echoes. For cases where target and clutter DoAs are closely spaced, we further introduce a joint spatial–Doppler domain clutter suppression scheme to achieve more comprehensive sensing and correct potential estimation biases or errors. Simulations demonstrate that CLAM-enabled ZF effectively suppresses clutter signals. Compared with conventional FFT/MUSIC and Doppler-only MTI baselines, the proposed methods achieve higher accuracy in estimating delays and DoAs for slow, weak, or near-angle targets in low-altitude environments, while incurring only moderate additional online complexity.






\bibliographystyle{IEEEtran}
\bibliography{IEEEabrv, references}

\end{document}